\documentclass[12pt]{article}
\usepackage{epsfig}
\usepackage{amssymb}
\usepackage{graphicx}
\usepackage{amsmath}
\usepackage{amssymb}
\usepackage{amsfonts}
\usepackage{bbm} 
\usepackage{subfigure}
\usepackage{color}
\def\eeq{\end{eqnarray}}

\def\p{\partial}

\def\D{\mathcal{D}}

\def\de{\partial}

\def\=:{=\hspace{-.7em}\raisebox{1.1ex}{.}\hspace{.1em}\raisebox{-0.2ex}{.} }

\newcommand{\beqn}{\begin{eqnarray}}
\newcommand{\eeqn}{\end{eqnarray}}

\newcommand {\beq}{\begin{eqnarray}}
\newcommand {\eeqq}{\end{eqnarray}}

\newcommand {\Tr}{{\rm Tr}\,}

\newcommand {\Nf}{N_{\rm f}}
\newcommand {\Nc}{N_{\rm c}}

\newcommand {\mc}{{\mathbb C}}

\def\changed#1{ #1}

\makeatletter
\@addtoreset{equation}{section}

\renewcommand{\thefootnote}{\fnsymbol{footnote}}

\makeatother

\begin{document}

\thispagestyle{empty}
\begin{flushright}
IFUP-TH/2008-35\\
October, 2008 \\
\end{flushright}
\vspace{3mm}

\begin{center}
{\Large \bf On the Stability of Non-Abelian Semi-local Vortices 
}
\\[12mm]
\vspace{5mm}

\normalsize { { Roberto~Auzzi$^{1}$, Minoru~Eto$^{2,3}$, 
Sven~Bjarke~Gudnason$^{2,3}$, \\
Kenichi~Konishi$^{2,3}$ and Walter~Vinci$^{2,3}$}}

\vskip 1.5em

{\small
$^1$ {\it University of Wales Swansea, 
Deptartment of Physics,
Singleton Park,   \\ 
Swansea, SA2 8PP,  Wales, UK}
\\
$^2$ {\it INFN, Sezione di Pisa,  Largo Pontecorvo, 3, Ed. C, 56127
 Pisa, Italy } 
\\
$^3$ {\it Department of Physics, University of Pisa,  Largo
 Pontecorvo, 3, Ed. C, 56127 Pisa, Italy } } 

\vspace{15mm}

\begin{abstract}
We study the stability of non-Abelian semi-local vortices based on an
$\mathcal{N}=2$ supersymmetric 
$H= SU(\Nc)\times U(1)/{\mathbb Z}_{\Nc}\sim U(\Nc) $ 
gauge theory with an arbitrary number of flavors ($N_f > N_c$) in the fundamental
representation, when certain  $\mathcal N=1$ mass terms are present,
making the vortex solutions no longer BPS-saturated.
Local (ANO-like) vortices are found to be stable against fluctuations
in the transverse directions. 
Strong evidence is found that the ANO-like vortices are
actually the true minima.  
In other words, the semi-local moduli, which are present in the BPS limit,
disappear in our non-BPS system, leaving the vortex with the
orientational moduli $\mc P^{\Nc-1}$ only.
We discuss the implications of this fact on the system in which the
$U(\Nc)$ model arises as the low-energy approximation of an underlying
e.g.~$G=SU(\Nc+1)$ gauge theory. 
\end{abstract}
\end{center}
\vfill
\newpage
\setcounter{page}{1} \setcounter{footnote}{0}
\renewcommand{\thefootnote}{\fnsymbol{footnote}}

\section{Introduction}
The last several years have witnessed a remarkable progress in our
understanding of physics of solitons, in particular, that of vortices 
in non-Abelian gauge theories. These problems are important as they
are intimately related to one of the deepest issues of particle
physics such as confinement \cite{'tHooft:1981ht}; they may be
important in some condensed-matter physics or cosmological problem as 
well. The progress has been made mainly in the context of
supersymmetric gauge theories where many aspects of calculations such 
as the quantum modification of the potential and symmetry
realizations, are under much better control. After the discovery of
vortices carrying continuous, non-Abelian moduli in \cite{HT,ABEKY}, 
many related studies have been carried out 
\cite{ABEK,tmon,SY,HT2,GSY,Duality,Baptista:2008ex}.    
For good reviews, see
Refs.~\cite{Eto:2006pg,Shifman:2007ce,Tong:2005un,Tong:2008qd}. 
Some recent developments involve the following issues: 

\begin{itemize}
\item A detailed study of moduli and the transformation properties of
 higher winding vortices \cite{HashiTong,Eto:2005yh,ASY,sevenHW} has
 been performed. The last of these papers contains basic results on
 the vortex transformation properties which allow them to be
 interpreted in a simple group-theoretical language. 

\item Another development concerns semi-local vortices in systems with
 larger number of flavors (matter fields) \cite{HT,SYSemi,SemiL}. 
 As it happens in the $U(1)$ Higgs systems with more than one charged
 scalar field \cite{Vachaspati:1991dz,achurev}, a new kind of moduli
 emerges; the thickness of the flux tube is no longer a fixed
 quantity, but corresponds to one of these moduli.
 This makes the corresponding vortex moduli space very interesting. 
 In particular, a new type of (Seiberg-like) duality has been found,
 between different models having dual vortex moduli (and sharing the
 common sigma-model lump limit at very strong couplings)
 \cite{SemiL}. 

\item A systematic study of non-BPS local  vortices has been initiated in
 Refs.~\cite{AEV1,AEV2} where several kinds of perturbations have been
 considered. The interactions have a very rich structure, depending
 both on the distance and relative orientations, a new feature not
 present in the Abelian counterpart.  

\item A particularly significant development concerns the extension of
 the analysis to systems with generic gauge groups
 \cite{KF,GFK,AGG}. As compared to the $U(N)$ models studied in most
 papers, the systems based on, e.g.   $U(1)\times SO(N)$, $U(1)\times
 USp(2N)$ are characterized by a larger vacuum degeneracy, even after
 the  ``color-flavor locked'' vacuum has been chosen to study the
 solitons. 

\item The question of dynamical Abelianization has recently been
 addressed more carefully in Ref.~\cite{KonishiMN}. Only under
 definite conditions, the non-Abelian moduli fluctuating along the
 vortex length and in time (which are described by a two-dimensional 
 sigma-model), are absorbed in the monopoles, turn into the dual
 gauge group, before Abelianizing dynamically.
 The recent result on the non-Abelian vortices with a product moduli
 \cite{DKO} seems to be particularly significant in this context.  

\item Further important developments have been achieved in the study of
 BPS non-Abelian vortices in theories with $\mathcal{N}=1$
 supersymmetry (heterotic vortices) \cite{heterotic,Tong:2008qd}. These works
 generalize the known links between the vortex and the bulk theory to
 a less supersymmetric case. It is reasonable to think that heterotic
 vortices play a role in the physics of Seiberg duality
 \cite{Seiberg:1994bz}. 
 However, so far, there seems to be no clean statement about this.
 Some related works can be found in Refs.~\cite{sdvort}. 

\item Another direction involves the non-Abelian vortex in the
 Higgs vacuum of $\mathcal{N}=1^*$ theory with gauge group $SU(N_c)$,
 which is a mass deformation of the conformal $\mathcal{N}=4$
 theory. This issue has been first studied for $N_c=2$ in
 Ref.~\cite{n1st1}. In Ref.~\cite{n1st2} the case of larger $N_c$ has
 been studied, both in the weakly coupled field theory and in the IIB
 string  dual (the Polchinski-Strassler background \cite{ps}). 

\end{itemize}

The present investigation is a natural extension of these lines of
research, in particular along the second and third. Namely, we study
the stability of the semi-local vortices in the presence of an 
${\cal{N}}=1$ perturbation which makes the vortices non-BPS. 
In the case of a theory with a mass gap, the 't Hooft standard
classification of possible massive phases \cite{thooft} is applicable;  
as a result, there is a very clear relation between the Higgs
mechanism and magnetic confinement. 
In the model discussed in this paper, this is not a priori clear due
to the presence of massless degrees of freedom (see Ref.~\cite{penin}
for some of the subtleties associated with the massless case). 

The question is important from the point of view that the vortex
system (with gauge group $H$) being studied is actually a low-energy
approximation of an underlying system with a larger gauge group, $G$.  
The homotopy-map argument shows that the regular monopoles 
arising from the symmetry breaking $G \to H$ are actually unstable in
the full theory, when the low-energy VEVs breaking completely the
gauge group are taken into account.  
In other words, such monopoles are actually confined by the vortices
developed in the low-energy $H$ system: this allows us to relate the
continuous moduli and group transformation properties of the vortices
to those of the monopoles appearing at the ends, thus explaining the
origin of the dual gauge groups, such as the
Goddard-Nuyts-Olive-Weinberg (GNOW) duals \cite{Goddard:1976qe}.  
The stability problem on non-BPS non-Abelian strings have also been
studied in another system: the Seiberg-dual theory of the ${\cal N}=1$
supersymmetric $SO(\Nc)$ QCD \cite{Eto:2007hf}. 

In these considerations a subtle but crucial point is that when small
terms arising from the symmetry breaking $G \to H$ are taken into
account, neither the high energy system (describing the symmetry
breaking $G \to H$ and regular monopoles) nor the low-energy system
(in the Higgs phase $H \to {\mathbf 1}$ describing the vortices) is
BPS-saturated any longer.   
This on the one hand allows the monopoles and vortices to be related
in a one-to-one map (each vortex ends up on a monopole); on the other
hand, the system is no longer BPS and we must carefully check the fate
of the moduli of the BPS vortices which do not survive the
perturbation.  In fact, zero modes related to global symmetries
(orientational modes) still survive, because of their origin as
Goldstone bosons, while other modes are no more protected by
supersymmetry. 
These issues are the subject for investigation in this paper.

\section{Stability of Non-Abelian Semi-local Vortices}

\subsection{Semi-local Vortices}

The question of stability of the semi-local vortices
\cite{Vachaspati:1991dz,achurev}   
for Abelian systems has been investigated by Hindmarsh
\cite{Hindmarsh:1991jq} and other authors \cite{Vachasp,leese}, some
time ago. 
The model which has been studied is an Abelian Higgs system with more
than one flavor $(\Nf>1)$, 
\beq \mathcal{L} =  -\frac{1}{4e^2}F_{\mu\nu}F^{\mu \nu} 
+ \D_\mu\phi (\D^\mu\phi)^\dag 
- \frac{\lambda}{2}\left(\phi\phi^\dag - \xi \right)^2 \ , 
\label{AbelianSL}\eeq
where $\D_{\mu }= \de_{\mu} - i A_{\mu}$ is the standard covariant
derivative, $\phi = (\phi_{1}, \phi_{2}, \ldots, \phi_{\Nf})$
represents a set of complex scalar matter fields of the same charge. 
This model is sometimes called the {\it semi-local} model since not
all global symmetries, i.e.~$U(\Nf)$ here, are gauged. As a
consequence, the vacuum manifold ${\cal M}$ is 
$\mc P^{\Nf-1} = SU(\Nf)/[SU(\Nf-1)\times U(1)]$. 
Since the first homotopy group of ${\cal M}$ is trivial for $\Nf>1$,
vortex solutions are not necessarily stable. 
For $\beta \equiv \lambda/e^{2} < 1$ (i.e.~type I superconductors) 
the vortex of ANO-type \cite{Abrikosov:1956sx,Nielsen:1973cs}  
is found to be stable.  In the interesting special (BPS) case,
$\beta=1$, there is a family of vortex solutions with the same
tension, $T=2\pi\xi$. Except for the special values of the
moduli (in the space of solutions),  which represent the ANO vortex
(sometimes called a ``local vortex''), the vortex has a power-like
tail in the profile function, hence the vortex width (thickness of the
string) can be of an arbitrary size\footnote{This type of vortex
 solutions has been termed  ``semi-local vortices''. Though it is not
 an entirely adequate term, we shall stick to it, as it is commonly
 used in the literature.}.
In the limit of large size, the vortex essentially reduces to
the  $\mathbb{C}P^{N_{f}-1}$ sigma-model lump
(or two-dimensional skyrmion), 
characterized by  $\pi_{2}(\mathbb CP^{N_{f}-1}) = {\mathbb Z}$. 

For $\beta > 1$ (i.e.~type II superconductors), vortices are found to
be unstable against fluctuations of the extra fields (flavors) which
increase the size (and spread out the flux).  

Properties of non-BPS solutions in the case of non-Abelian vortices,
within the models similar to Eq.~(\ref{AbelianSL}) but with a $U(\Nc)$
gauge group, have been studied by some of us \cite{AEV2}.  
The interactions among the vortices are found to depend on the
relative orientations carried by these non-Abelian vortices as well as
the distances between the vortices. 
The two new regimes\footnote{At large distance, the type
   I$^{*}$ interaction is attractive for parallel vortices and
   repulsive for anti-parallel vortices. Vice versa for type II$^{*}$
   interaction. } called type I$^*$/II$^*$ have been found 
in addition to the usual type I/II superconductors~\cite{AEV2}.

The same authors investigated furthermore another class of (type
I/I$^*$) systems in a super Yang-Mills theory as well \cite{AEV1},
characterized by certain non-vanishing adjoint scalar masses.  
These models are potentially important as they are exactly the sort of
systems arising as the low-energy approximation as a result of
symmetry breaking at some higher mass scale.
Under such circumstances, the properties of the vortices in the
low-energy system are closely related to those of the regular
monopoles arising at high energies. With this motivation in mind,   
we shall here concentrate on this class of non-BPS vortices.

\subsection{The Model and the Vortex Solution\label{model}}

The model we consider is an $\mathcal{N}=2$ supersymmetric $(U(\Nc)$ 
gauge theory with $\Nf>\Nc$ hypermultiplets (quarks) 
$Q_f,\tilde Q_f$ $(f=1,\dots,\Nf)$.
Since all the essential features are already present in the minimal case
i.e.~$\Nc=2$, we will concentrate  in the
following on this case, i.e.,  the $U(2)$ model\footnote{
The extension to more general cases, $\Nc>2$, will be discussed in
 appendix \ref{app}.}.
The superpotential at hand is
\beq W = \frac{1}{\sqrt{2}}
\left[\tilde{Q}_f (a_0 + a_i \tau^i+\sqrt{2} m_f) Q_f 
+ W_e(a_0)+ W_g(a_i \tau^i)\right]\ , \eeq 
where 
\beq W_e= - \xi \,a_0 + \eta  \, a_0^2\ , \quad 
W_g(a_i\tau^i)=\mu \, a_i a_i\ . \eeq
Two real positive mass parameters $\eta $ and $\mu $ have been
introduced for the adjoint scalars. 
The terms proportional  to $\eta,\mu$  break ${\cal N}=2$
supersymmetry (SUSY) to ${\cal N}=1$.
$m_f$ are the (bare) quark masses.
$\xi$ is the $F$-term Fayet-Iliopoulos (FI) parameter. 
In fact, it is $SU(2)_{R}$-equivalent to the standard FI term. If only
the  $\xi$ term is kept, the system remains ${\cal N} =2$ supersymmetric. 

This kind of system naturally arises in the $\mathcal{N}=2$, $SU(3)$
SQCD, with SUSY softly broken down to $\mathcal{N}=1$ with a mass term
of the form  
\beq W = \kappa \, \textrm{Tr} \Phi^2 \ .  \eeq
Indeed, when the bare masses for the squarks are tuned to some special 
values, there exist quantum vacua in which the non-Abelian gauge
symmetry $SU(2)\times U(1)$ is preserved (the so-called $r=2$ vacua)
\cite{CKM}.    
The low energy effective theory in these vacua is exactly the theory
we are studying here.    
As the quantum vacua with an $SU(2)$ magnetic gauge group require the
presence of a sufficient number of flavors, $N_{f} \ge 2\,r = 4$, we
must necessarily deal with a system with an excess number of flavors
($N_{f} > N_{c}$).   
As these systems in the BPS approximation ($\eta,\mu=0$) contain
semi-local vortices with arbitrarily large widths, the
necessity of studying the fate of these vortices in the presence of
the perturbations $\eta,\mu$ presents itself quite naturally. 

The bosonic part of the Lagrangian 
is (we use the same symbols for the scalars as for the corresponding 
superfields):
\begin{align}
\mathcal{L} =&\ 
-\frac{1}{4 g^2} (F_{\mu \nu}^i)^2 - \frac{1}{4e^2}(F_{\mu \nu}^0)^2+
\frac{1}{g^2} |\D_{\mu} a_i|^2+ \frac{1}{e ^2}|\partial_{\mu} a_0|^2
\nonumber\\
&+(\D_\mu Q_f)^{\dagger}\D^\mu Q_f 
+ \D_\mu \tilde{Q}_f (\D^\mu \tilde{Q}_f)^{\dagger}
- V(Q,\tilde{Q},a_i,a_0)\ , \label{azione-tutta}
\end{align}
where $e $ is the $U(1)$ gauge coupling and $g$ is the $SU(2)$ gauge coupling.
The covariant derivatives and field strengths, respectively, are
defined by 
\begin{align}
&\D_\mu \left(Q_f,\tilde Q_f^*\right) 
=\left(\p_\mu - i A_\mu^i\frac{\tau^i}{2} -\frac{i}{2}A_\mu^0\right) 
\left(Q_f,\tilde Q_f^*\right) \ , \quad
\D_\mu a_i = \de_\mu a_i + \epsilon^{ijk}A_\mu^j a _k \ , \\
&F_{\mu\nu}^i = \de_\mu A_\nu^i -\de_\nu A_\mu^i + 
\epsilon^{ijk} A_\mu^j A_\nu^k \ , \quad 
F_{\mu\nu} = \de_\mu A_\nu^0 - \de_\nu A_\mu^0 \ . \nonumber
\end{align}
The potential $V$ is the sum of the following $D$ and $F$ terms
\begin{align}
V &=  V_1+V_2+V_3+V_4\ , \nonumber\\
V_1 &= \frac{g^2}{8}
\left( \frac{2}{g^2} \epsilon^{ijk} \bar{a}_{j} a_k +
\Tr_{\rm\! f} [Q^\dagger \tau^i Q] -
\Tr_{\rm\! f}[\tilde{Q} \tau^i \tilde{Q}^\dagger] \right)^2\ ,
\nonumber\\
V_2 &= \frac{e ^2}{8} \left(\Tr_{\rm\! f} 
[Q^\dagger Q]-\Tr_{\rm\! f}[\tilde{Q} \tilde{Q}^\dagger] \right)^2\ ,
\nonumber\\
V_3 &= \frac{g^2}{2} \left|\Tr_{\rm\! f} 
[\tilde{Q} \tau^i Q]+2 \, \mu  a_i \right|^2 +
\frac{e ^2}{2} \left|\Tr_{\rm\! f} 
[\tilde{Q}  Q]- \xi +2 \, \eta  a_0 \right|^2\ ,
\nonumber\\
V_4 &= \frac{1}{2} \sum_{f=1}^{\Nf} 
\left|(a_0+\tau^i a_i+\sqrt{2} m_f) Q_f \right|^2+
\frac{1}{2} \sum_{f=1}^{\Nf} 
\left|(a_0+\tau^i a_i+\sqrt{2} m_f) 
\tilde{Q}^{\dagger}_{f}\right|^2\ ,  \label{eq:potentialsep}
\end{align}
where $\Tr_{\rm\! f}$ denotes a trace over the flavor indices.
The squark multiplets are kept massless in the remainder of the paper
\beq m_f = 0\ . \eeq

The theory has a degenerate set of vacua in the Higgs phase where the
gauge symmetry is completely broken, which is the cotangent bundle
over the complex Grassmanian manifold 
$\mathcal M_{\rm vac}=Gr_{\Nf,\Nc} 
\simeq SU(\Nf)/[SU(\Nc)\times SU(\Nf-\Nc)\times U(1)]$ 
\cite{Lindstrom:1983rt} with $\Nc=2$. 
It is important to notice that the first homotopy group
of the Grassmanian manifold is trivial (for $\Nf> \Nc$).
Up to gauge and flavor rotations, we can choose the following VEV for
the scalar fields
\beq
Q=\tilde{Q}^\dagger=\sqrt{ \frac{\xi}{2} }\left(\begin{array}{ccccc}
1 & 0 & 0 &\cdots & 0\\
0 & 1 & 0 &\cdots & 0 \\
\end{array}\right) \ , \quad a_0 = 0 \ , \quad  a = 0 \ , 
\label{eq:vac} 
\eeq
where $a \equiv a_i \tau^i/2$.
The vacuum is  invariant under the following global color-flavor
locked rotations 
($U_{\rm c} \in SU(2)$, $U_{\rm f} \in SU(2) \subset SU(\Nf)$)\footnote{
The vacuum is also invariant under pure $SU(\Nf-2)$ flavor rotations
which act on the last $\Nf-2$ columns.}:
\beq Q \rightarrow U_{\rm c} Q  U_{\rm f}^\dagger \ , \quad 
\tilde{Q} \rightarrow U_{\rm f}^\dagger \tilde{Q}  U_{\rm c} \ , \quad
a \rightarrow U_{\rm c} a U_{\rm c}^\dagger \ , \quad 
F_{\mu \nu} \rightarrow U_{\rm c} F_{\mu \nu} U_{\rm c}^\dagger \ .
\label{colflavsymm}
\eeq
For $\eta\neq 0$, the theory has also another classical vacuum in the
Coulomb phase
\beq Q=\tilde{Q}^{\dagger}=0 \ , \quad a_0=\frac{\xi}{2\eta} \ , \quad a=0\ , 
\eeq
which ``runs away'' to infinity for $\eta =0$. 
In what follows, we consider the vacuum (\ref{eq:vac}) and the
non-Abelian vortices therein.

Next, we will construct a particular solution for the fundamental
(i.e.~the minimum winding) vortex, simply by embedding the well-known
solution for $\Nf=2$ flavors in our model.  
In the next section the stability of this solution under perturbations
of the additional flavors will be studied.  
Setting $\tilde{Q}=Q^\dagger$, the Euler-Lagrange equations for the
theory are 
\beq
\nonumber \partial_\mu F^{\mu \nu}_0 &=& - i e ^2
\left( Q_f^\dagger \D^\nu Q_f - (\D^\nu Q_f)^\dagger Q_f  \right)\ , \\
\nonumber  
\D_\mu F^{\mu\nu}_i&=& - i g^2
\left( Q_f^\dagger \tau_i \D^\nu Q_f - (\D^\nu Q_f)^\dagger \tau_i Q_f  \right) 
- \epsilon_{ijk} \left(a_j (\D^\nu a_k)^\dagger  + \bar{a}_j \D^\nu
a_k \right) , \\ 
\D^\mu \D_\mu Q&=&-\frac{1}{2}\frac{\delta V}{\delta Q^\dagger}\ , \qquad
\partial^\mu \partial_\mu a_0 =- e^2 \frac{\delta V}{\delta \bar{a}_0}\ ,\qquad
\D^\mu \D_\mu a_i = -g^2 \frac{\delta V}{\delta \bar{a}_i}\ .
\label{feq}
\eeq
We start from the standard Ansatz for the local vortex embedded in the model
with additional flavors
\beq
Q =\left(\begin{array}{ccccc}
\phi_0(r) e^{ i \theta} & 0 & 0 & \dots &0\\
0 & \phi_1(r) & 0& \dots & 0 \\
\end{array}\right)\ ,\quad 
a_0 = \lambda_0(r)\ ,\quad 
a = \lambda_1(r)\frac{\tau^3}{2}\ ,\nonumber\\
A_i =  A_i^\alpha \frac{\tau^\alpha}{2} 
= - \frac{\epsilon_{ij} x_j}{r^2} \left[1-f_1(r)\right] 
\frac{\tau^3}{2}\ , \quad 
A_i^0 = -\frac{\epsilon_{ij} x_j}{r^2} \left[1-f_0(r)\right]\ , 
\qquad\quad \label{ansa}
\eeq
with $r^2 = x_1^2+x_2^2$.
Notice that the adjoint fields $a_0,a=a_i\tau^i/2$ are non-trivial
when we consider the non-BPS corrections, whereas they vanish
everywhere if $\eta,\mu$ are set to zero (i.e.~the BPS limit).

\begin{figure}[h!tp]
\begin{center}
\includegraphics[width=0.5\linewidth]{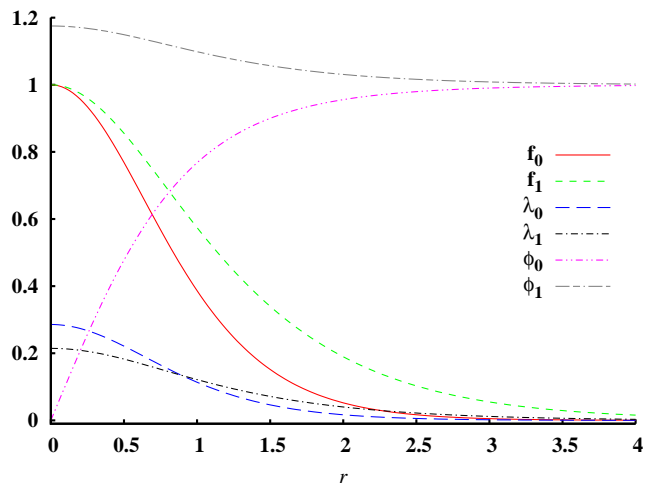}
\caption{\footnotesize 
Profile functions for the vortex in the radial direction $r$: $f_0$
(unbroken/red), $f_1$ (short-dashed/green), $\lambda_0$
(long-dashed/blue), $\lambda_1$ (dash-dotted/black),
$\phi_0$ (dash-double-dotted/magenta), $\phi_1$ (long-dash
short-dash-dotted/grey). The numerical values are taken as
$\xi=2,e=2,g=1,\eta =0.2, \mu =0.4$. For this configuration
$T=0.964 T_{\rm BPS}$. }
\label{profili} 
\end{center}
\end{figure}

By plugging the ansatz (\ref{ansa}) into Eq.~(\ref{feq}), we get a set
of complicated second order differential equations for the six
functions $\{\phi_0,\phi_1,f_0,f_1,\lambda_0,\lambda_1\}$
\cite{AEV1}. 
It is summarized in Appendix \ref{app}, see 
Eqs.~(\ref{eq:eom_01})-(\ref{eq:eom_06}) with $\Nc=2$.
This expression must be solved with the appropriate boundary conditions
\begin{eqnarray}
\nonumber f_1(0) = 1,\quad f_0(0)=1, \quad f_1(\infty)=0, \quad   f_0(\infty) = 0, \\
\phi_0(\infty) = 1,\quad \phi_1(\infty) = 1, \quad
\lambda_0(\infty) = 0, \quad   \lambda_1(\infty) = 0.
\end{eqnarray}
We find the following behavior for small $r$
\beq
\phi_0 \propto {\cal O}(r), \,\,\, \phi_1 \propto {\cal O}(1), \,\,\, \lambda_0
\propto {\cal O}(1), \,\,\, \lambda_1 \propto {\cal O}(1). 
\eeq
A numerical solution is shown in Fig.~\ref{profili}.

Further solutions can be obtained by acting with the $SU(2)_{\rm c+f}$
symmetry (\ref{colflavsymm}) on this Ansatz.  
As a consequence it develops some internal (orientational) zero modes. 
In fact, the vortex leaves a $U(1)_{\rm c+f}$ subgroup of 
$SU(2)_{\rm c+f}$ unbroken, i.e.~these zero modes parameterize the
space $\mathbb{C}P^{1}=SU(2)/U(1)=S^2$.

In the BPS limit $\eta=\mu=0$, the system has vortices which are
solutions to the first-order equations \cite{ABEKY}, see also
Eqs.~(\ref{eq:bps01}) and (\ref{eq:bps02}), with tension  
\beq T_{\rm BPS} =  2 \pi\xi \ . \eeq
As $\lambda_{0}(x)=\lambda_1(x)=0$ ($a_{0}= a =0$) in this BPS
configuration, its substitution into the  Lagrangian
(\ref{azione-tutta}) gives the BPS value, 
$T_{\rm BPS} =  2 \pi \xi$, 
even if it is no longer a solution to the vortex equations
(\ref{feq}), for $\eta,\mu \ne 0$. 
This means that the non-BPS vortex tension derived from
Eq.~(\ref{feq}) is necessarily {\it less} than the BPS value. 
In other words, the model we are considering always yields type I
superconductivity \cite{AEV1}. 

Our result below -- the stability of the ANO-like vortices  -- 
thus generalizes naturally the known correlation between the type of superconductor and the kinds of stable vortices, found in the Abelian Higgs models \cite{Hindmarsh:1991jq,Vachasp,leese}. 
Also, in so much as we study a softly-broken ${\cal N}=2$ supersymmetric gauge theory and its low-energy vortex system, the present work can be regarded as an extension of the one in $SU(2)$ gauge theory
\cite{Vainshtein:2000hu}, even though in the latter the low-energy system was naturally Abelian  ($U(1)$).  We have verified that the same qualitative conclusion holds in that case.

\subsection{Fluctuation Analysis \label{sec:zma}}

In this section, we will generalize the \changed{fluctuation} analysis of the
stability of the Abelian vortex of Ref.~\cite{Hindmarsh:1991jq} to the
non-Abelian vortex.
As we have mentioned, the first homotopy group of the vacuum manifold
is trivial, thus the local vortex found in the previous section can be
unstable. 
To study the stability of non-BPS local vortices, we must
consider the quadratic variations of the Lagrangian due to small
perturbations of the background fields. 
It turns out that the quadratic variation of the Lagrangian can be
written as the sum of two pieces
\beq
\delta^2 \mathcal{L} = \left.\delta^2\mathcal{L}\right|_{\rm local \ fields}
+ \left.\delta^2\mathcal{L}\right|_{\rm semi-local \ fields}\ , 
\label{Eq:decomp} \eeq
where the first term denotes the variation with respect to the fields
describing a non-trivial background vortex configuration  (i.e.~the
``local'' fields), while the second term is the variation with respect
to the ``semi-local'' fields 
($Q_f, \tilde{Q}_f$ with $\Nf\geq f > \Nc=2$).
The key point is that there are no mixed terms (at the second order)
between the variations of the background fields and the ``semi-local''
fields. The first term cannot give rise to instabilities (i.e.~the
local vortices with $\Nf = 2$ are topologically stable). 
Hence, it suffices for our purpose to study only the second term 
\beq
\left.\delta^2\mathcal{L}\right|_{\rm semi-local \ fields} = 
(\D_{{\rm b}\mu} \delta Q_3)^{\dagger} 
\D_{{\rm b}}^\mu\delta Q_3 +
(\D_{{\rm b}\mu} \delta \tilde{Q_3})
(\D_{{\rm b}}^\mu \delta \tilde{Q_3})^{\dagger} - 
\left.\delta ^2 V\right|_{Q_3,\tilde{Q}_3}\ ,\label{Eq:variation} 
\eeq
where  we have taken, for simplicity and without loss of generality,
$\Nf=3$. The subscript `${\rm b}$' denotes background fields which we
fix to be given by the Ansatz (\ref{ansa}) in the following.
Let us work out this variation explicitly. 
Even if $\tilde{Q_{\rm b}}^{\dagger}=Q_{\rm b}$  in the background
fields we keep their variations independent.

Let us first consider the variation of the potential
(\ref{eq:potentialsep}), piece by piece:

\subsubsection*{Variation of $V_1$}

Remember that we are only interested in the variations that involve
the semi-local fields, i.e.~the variation with respect to the third
flavor. The crucial observation is that the background value for this
field is zero: $\tilde{Q}_{3{\rm b}}^{\dagger}=Q_{3{\rm b}}=0$.  
This means that the variation of terms such as 
$\Tr_{\!\rm f}[Q^\dagger \tau^i Q]$ are already quadratic
in the perturbation. 
Furthermore, the term $\epsilon^{ijk} \bar{a}_{j} a_k$ evaluated on
the background fields is zero, due to $a_3$ being the only
non-zero non-Abelian adjoint field. 
Thus, the variation of $V_1$ with respect to the semi-local fields is
at least cubic, coming from the cross terms involving the adjoint
field. Hence, the quadratic variation vanishes
\beq \left.\delta^2 V_1\right|_{Q_3,\tilde Q_3} = 0 \ . \eeq

\subsubsection*{Variation of $V_2$}

$V_2$ contains  no adjoint fields. The same argument used for $V_1$  
goes through: the variations of $V_2$ are at least quartic
\beq \left.\delta^2 V_2\right|_{Q_3,\tilde Q_3} = 0 \ . \eeq

\subsubsection*{Variation of $V_3$}

The variation of $V_3$ gives us the first non-trivial contribution. 
In this case the variation with respect to the semi-local fields is at
least quadratic, and it is given by 
\begin{align}
\left.\delta^2 V_3\right|_{Q_3,\tilde Q_3} &= \frac{g^2}{2} 
\left(\Tr_{\rm \!f} (Q_{\rm b}^\dagger \tau^3 Q_{\rm b})
+2 \, \mu  a_{3{\rm b}} \right)
\left(\delta\tilde{Q}_3 \tau^3 \delta Q_3+{\rm c.c.}\right) \nonumber\\
&+ \frac{e ^2}{2} \left(\Tr_{\rm \!f} (Q_{\rm b}^\dagger  Q_{\rm b})-
\xi +2 \, \eta  a_{0{\rm b}} \right)
\left(\delta\tilde{Q}_3 \delta Q_3+{\rm c.c.}\right)\ .
\end{align}

\subsubsection*{Variation of $V_4$}

The variation of $V_4$ is also quadratic and is simply
\beq
\left.\delta^2 V_4\right|_{Q_3,\tilde Q_3} = 
\frac{1}{2}\delta Q_3^\dagger 
\left(a_{0{\rm b}}+\tau^3 a_{3{\rm b}}\right)^2 \delta Q_3
+\frac{1}{2}\delta \tilde{Q}_3 
\left(a_{0{\rm b}}+\tau^3 a_{3{\rm b}}\right)^2 
\delta\tilde{Q}_3^\dagger \ .
\eeq

The variations are not diagonal, thus involve mixed terms like 
$\delta\tilde Q_3 \delta Q_3$. 
We can easily diagonalize them by the following change of coordinates
(keeping the kinetic terms canonical)
\beq \delta Q_3 = \frac{1}{\sqrt{2}}(q+\tilde{q}^{\dagger})\ , \qquad 
\delta \tilde{Q}_3  =\frac{1}{\sqrt{2}}(q^{\dagger}-\tilde{q})\ ,
\label{diagtransf}\eeq
which yields the following variation of the potential to second order
\begin{align}
\left.\delta^2 V\right|_{Q_3,\tilde{Q}_3} &= \frac{g^2}{2} 
\left(\Tr_{\rm \!f} (Q_{{\rm b}}^\dagger \tau^3 Q_{\rm b})+2 \, 
\mu  a_{3{\rm b}}\right) 
\left( |q_I|^2 -|\tilde{q}_I|^2-|q_{II}|^2 +|\tilde{q}_{II}|^2 \right)
\nonumber\\
&+
\frac{e ^2}{2} \left(\Tr_{\rm \!f} (\tilde{Q}_{\rm b} Q_{\rm b})
-\xi+2 \, \eta  a_{0{\rm b}} \right) \left(|q_I|^2 -|\tilde{q}_I|^2+|q_{II}|^2 -|\tilde{q}_{II}|^2 \right)
\nonumber\\
&+
\frac{1}{2}\left(a_{0{\rm b}}+a_{3{\rm b}}\right)^2
\left( |q_I|^2 +|\tilde{q}_I|^2\right)
+\frac{1}{2}\left(a_{0{\rm b}}-a_{3{\rm b}}\right)^2\left( |q_{II}|^2
+|\tilde{q}_{II}|^2\right)\ , 
\end{align}
where the capital indices $I$ and $II$ label the color components. The
problem of stability is now reduced to studying four decoupled
Schr\"odinger equations. We expand the fluctuations as 
\beq q_{I,II}\equiv \sum_k \psi^{(k)}_{I,II} e^{i k \theta}\ , \quad 
\tilde{q}_{I,II} \equiv 
\sum_k \tilde{\psi}^{(k)}_{I,II} e^{-i k\theta} \ . \eeq
Using these expansions, the quadratic variations of the energy density 
(\ref{Eq:variation}) give rise to the following Schr\"odinger equations
\beq
\left[-\frac{1}{r}\frac{d}{dr}\left(r \frac{d}{dr} \right) 
+ V^{(k)}_{I,II}\right]\psi^{(k)}_{I,II} 
&=& M^{(k)}_{I,II} \psi^{(k)}_{I,II} \ ,   \nonumber\\
\left[-\frac{1}{r}\frac{d}{dr}\left(r \frac{d}{dr} \right) 
+ \tilde V^{(k)}_{I,II}\right]\tilde \psi^{(k)}_{I,II} 
&=& \tilde M^{(k)}_{I,II} \tilde\psi^{(k)}_{I,II} \ ,
\label{Schrodinger}
\eeq
\begin{figure}[h!tp]
\begin{center}
\mbox{
\subfigure[]{\resizebox{!}{0.24\linewidth}{\includegraphics{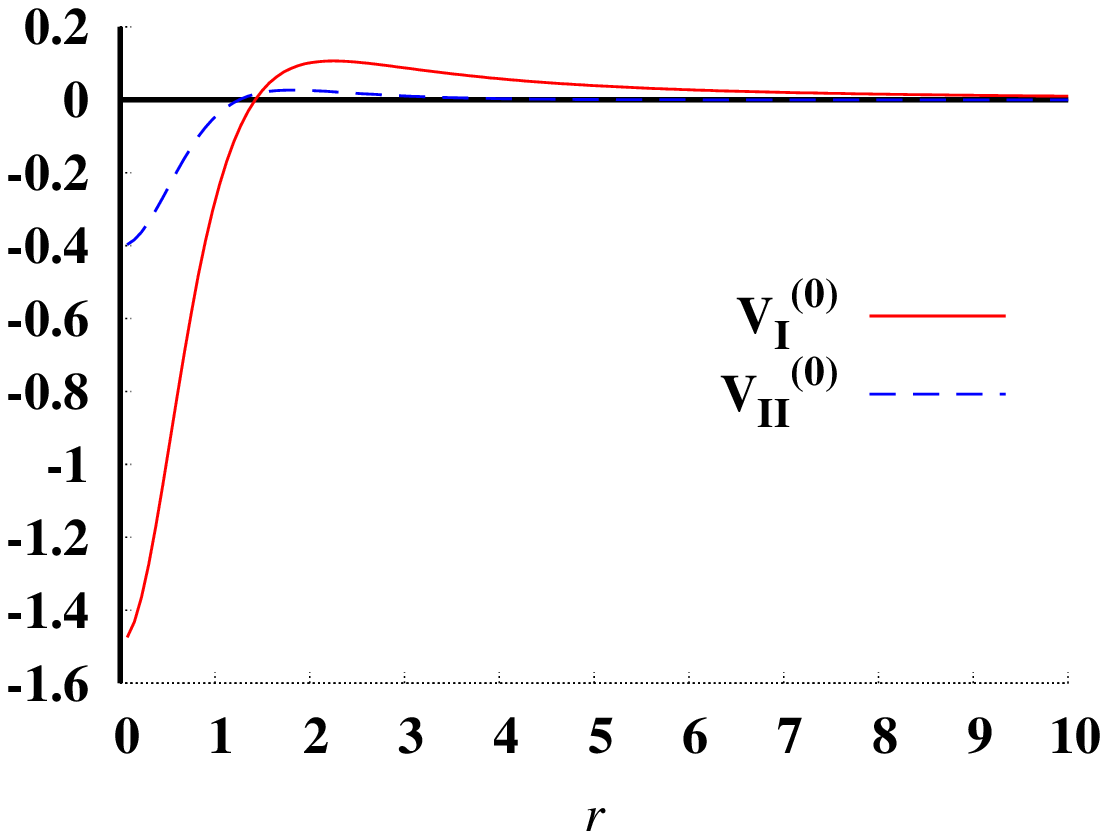}}}\quad
\subfigure[]{\resizebox{!}{0.24\linewidth}{\includegraphics{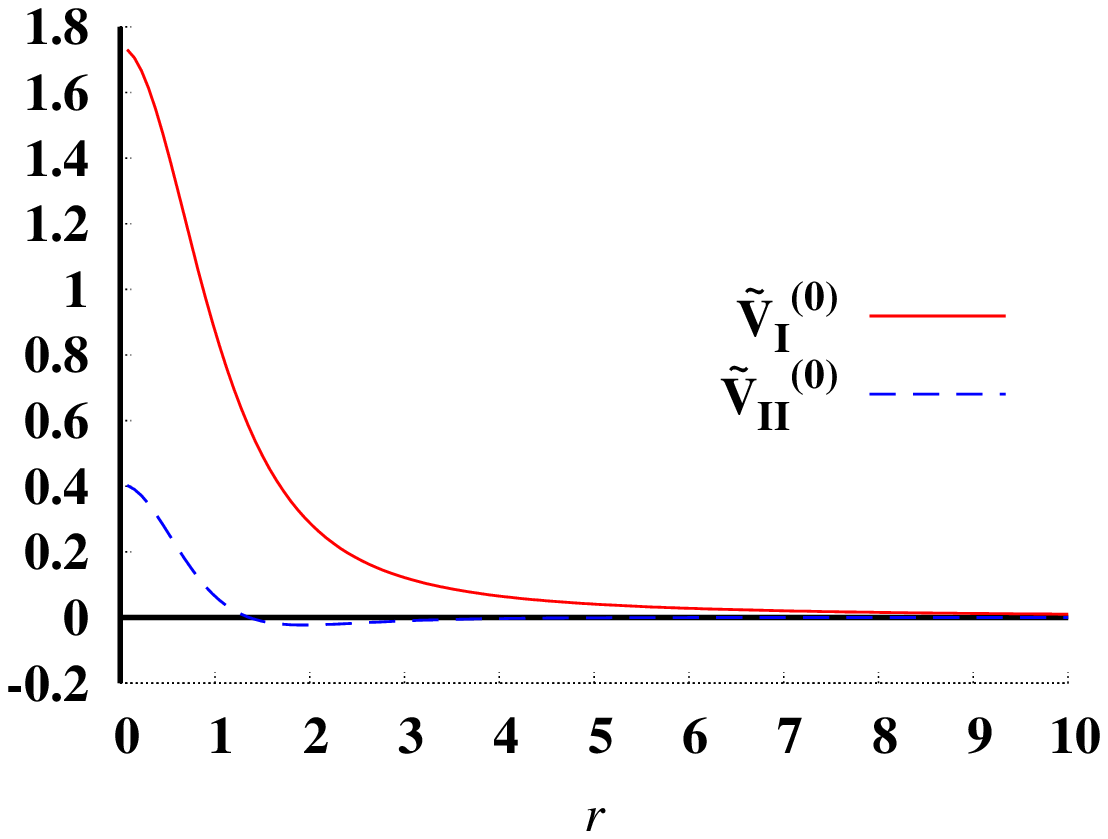}}}\quad
\subfigure[]{\resizebox{!}{0.24\linewidth}{\includegraphics{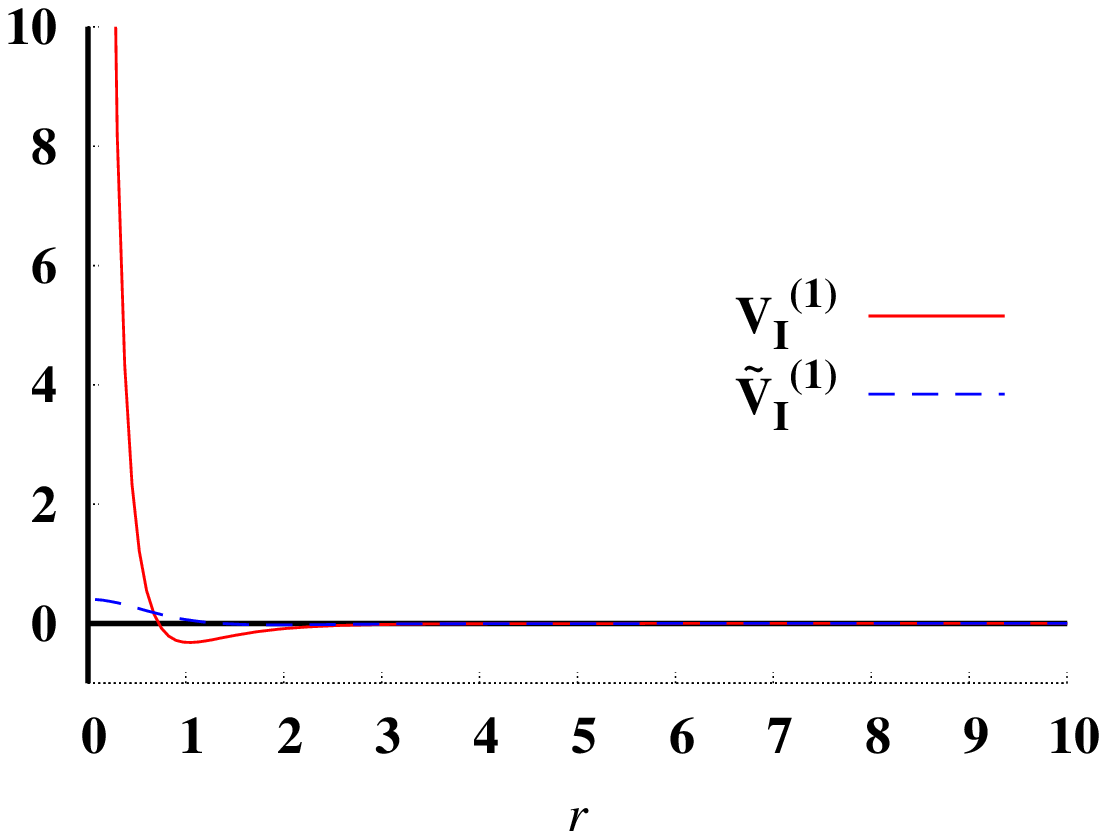}}}}
\end{center}
\vspace{-0.7cm}
\caption{\footnotesize Potentials as functions of $r$: 
(a) $V^{(0)}_I$ and $V^{(0)}_{II}$. 
(b) $\tilde V^{(0)}_I$ and $\tilde V^{(0)}_{II}$.
(c) $V^{(1)}_I$ and $\tilde V^{(1)}_I$.
The results are plotted for $e=2, g=1, \xi=2, \eta =0.2$ and
$\mu =0.4$.}  
\label{potentials} 
\end{figure}
which must be solved with the following boundary conditions at small $r$:
\beq
\psi^{(k)}_{I,II}(r),\tilde \psi^{(k)}_{I,II}(r)= r^{k}+\mathcal O(r^{k+1}).
\eeq
The effective potentials are
\begin{align}
V^{(k)}_{I,II} &= \frac{1}{4 r^2}\left[f_0-1\pm (f_1-1) + 2k\right]^2
+\frac{1}{2}(\lambda_0\pm \lambda_1)^2 \\
&+ \frac{e ^2}{2}(\phi_0^2+\phi_1^2-\xi+ 2 \, \eta  \lambda_0)
\pm \frac{g^2}{2}(\phi_0^2-\phi_1^2+ 2 \, \mu  \lambda_1) \nonumber\\
\tilde{V}^{(k)}_{I,II} &= \frac{1}{4r^2}\left[f_0-1\pm(f_1-1)+ 2k\right]^2
+\frac{1}{2}(\lambda_0\pm \lambda_1)^2 \\
&- \frac{e ^2}{2}(\phi_0^2+\phi_1^2-\xi+ 2 \, \eta \lambda_0) 
\mp \frac{g^2}{2}(\phi_0^2-\phi_1^2+ 2 \, \mu  \lambda_1) \ .
\end{align}
The upper signs refer to the color-index $I$, while the lower to index
$II$. 
Since only the first terms are dependent on $k$, it is easy to check
that
\begin{figure}[!bp]
\begin{center}
\mbox{
\subfigure{\resizebox{!}{0.3\linewidth}{\includegraphics{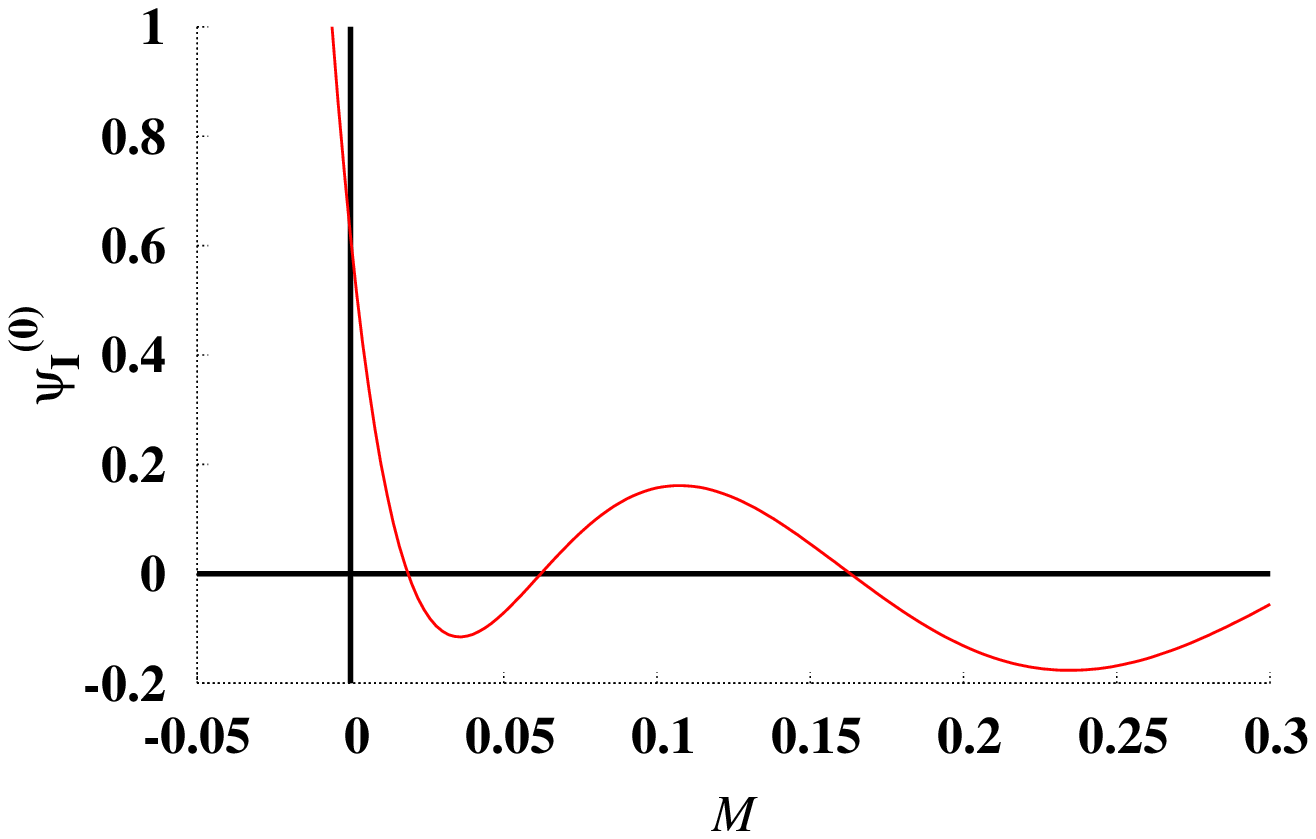}}}\quad
\subfigure{\resizebox{!}{0.3\linewidth}{\includegraphics{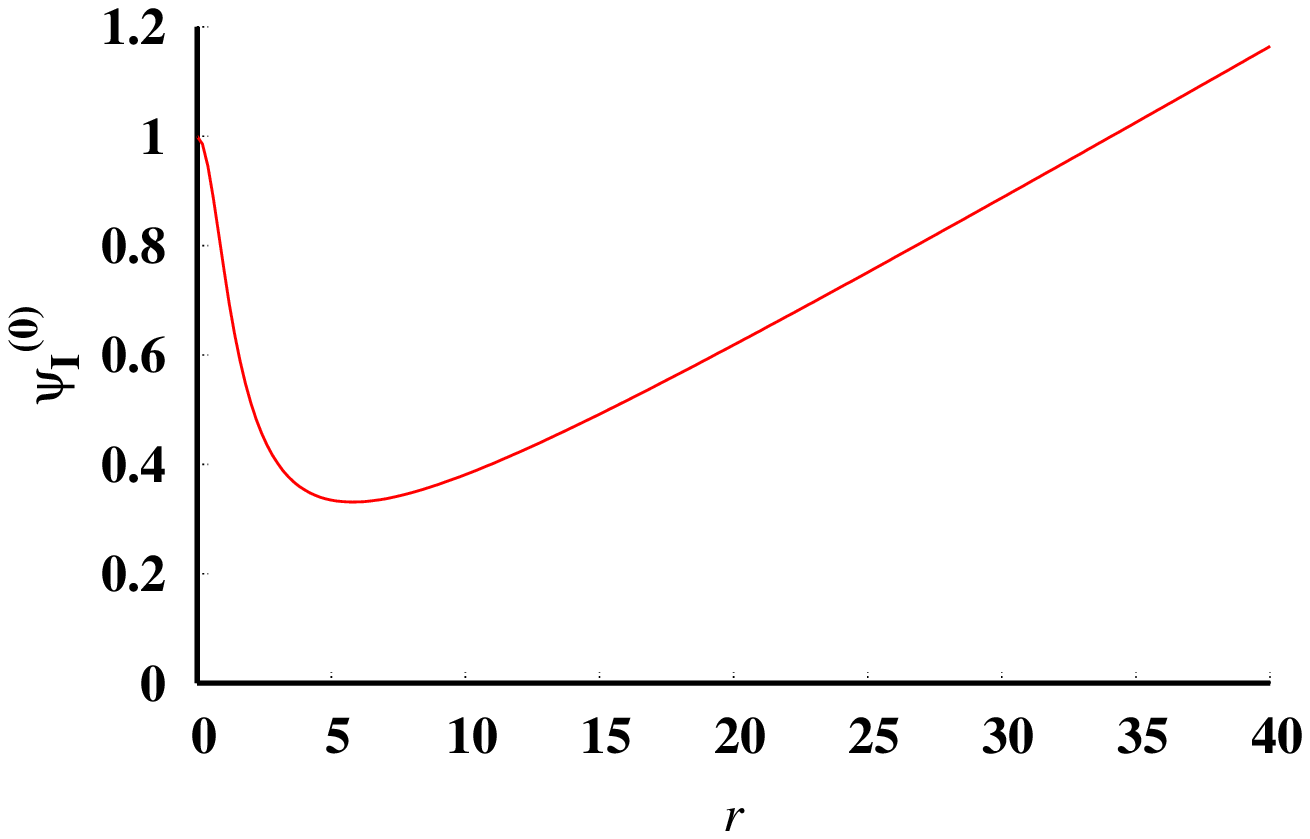}}}}
\end{center}
\vspace{-0.7cm}
\caption{\footnotesize Left panel: the {\it value} of the wave
 function $\psi_I^{(0)}$ at large distance ($r=20$, in the unit of
 the vortex size), as a function of $M$ (see
 Eqs. (\ref{Schrodinger})). There are no
 zeros at negative values of $M$: this implies the absence of
 normalizable tachyonic modes. 
 Right panel: the wave function for the zero energy fluctuation:
 $M=0$. The linear behavior is a common feature for zero energy wave
 functions. The numerical values of the parameters are chosen as in 
 Fig. \ref{potentials}.}  
\label{NBPSpotentials}
\end{figure}
\beq
\left(V^{(k)}_I,V^{(k)}_{II}\right) \ge 
\left(\min\left[V^{(0)}_I,V^{(1)}_I\right],V^{(0)}_{II}\right)\ ,
\quad 
\left(\tilde V^{(k)}_I,\tilde V^{(k)}_{II}\right) \ge 
\left(\min\left[\tilde V^{(0)}_I,\tilde V^{(1)}_I\right],
\tilde V^{(0)}_{II}\right)\ ,
\eeq
for all $k$.
To prove the above inequalities, it is sufficient to recall that the
profile functions for the gauge fields satisfy: $0\leq f_{0,1} \leq 1$.  

\begin{figure}[!tp]
\begin{center}
\mbox{
\subfigure[]{\resizebox{!}{0.2\linewidth}{\includegraphics{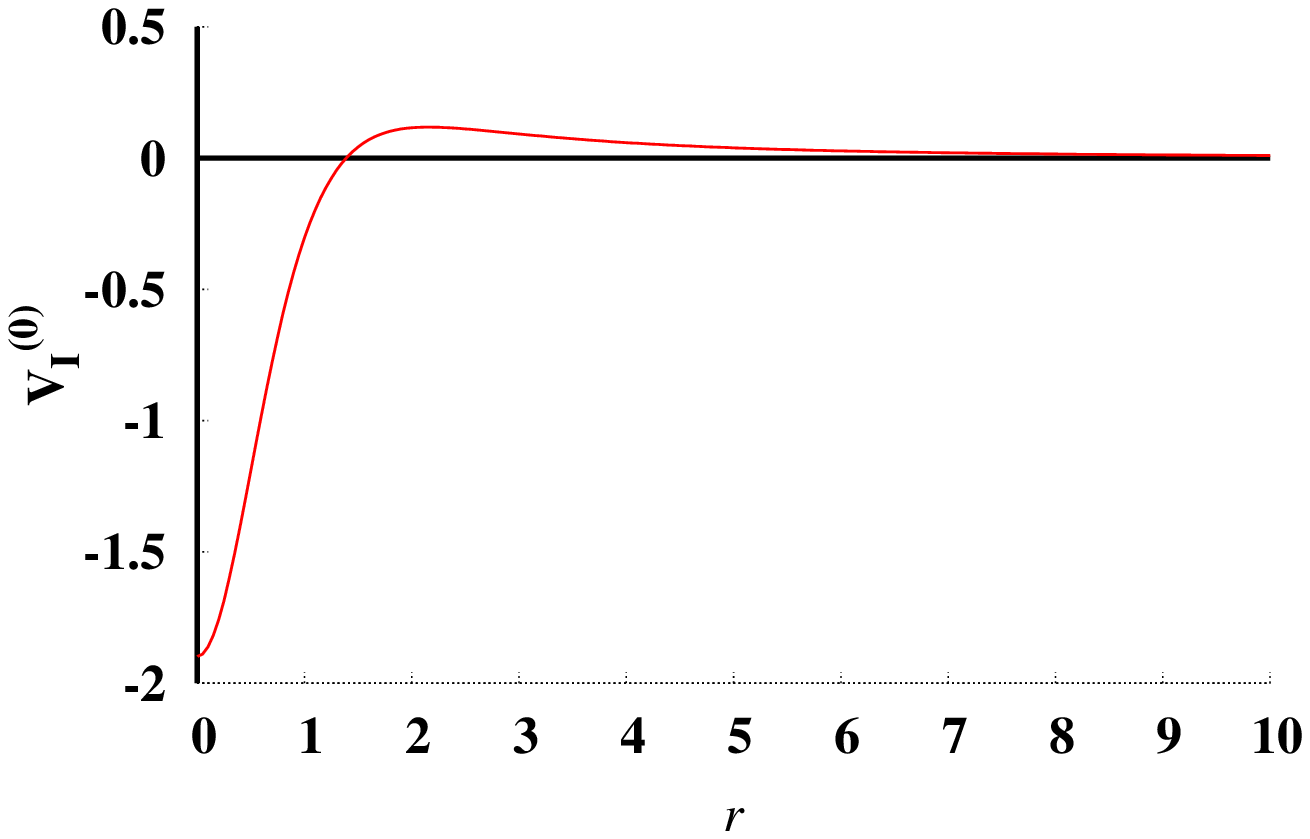}}}\quad
\subfigure[]{\resizebox{!}{0.2\linewidth}{\includegraphics{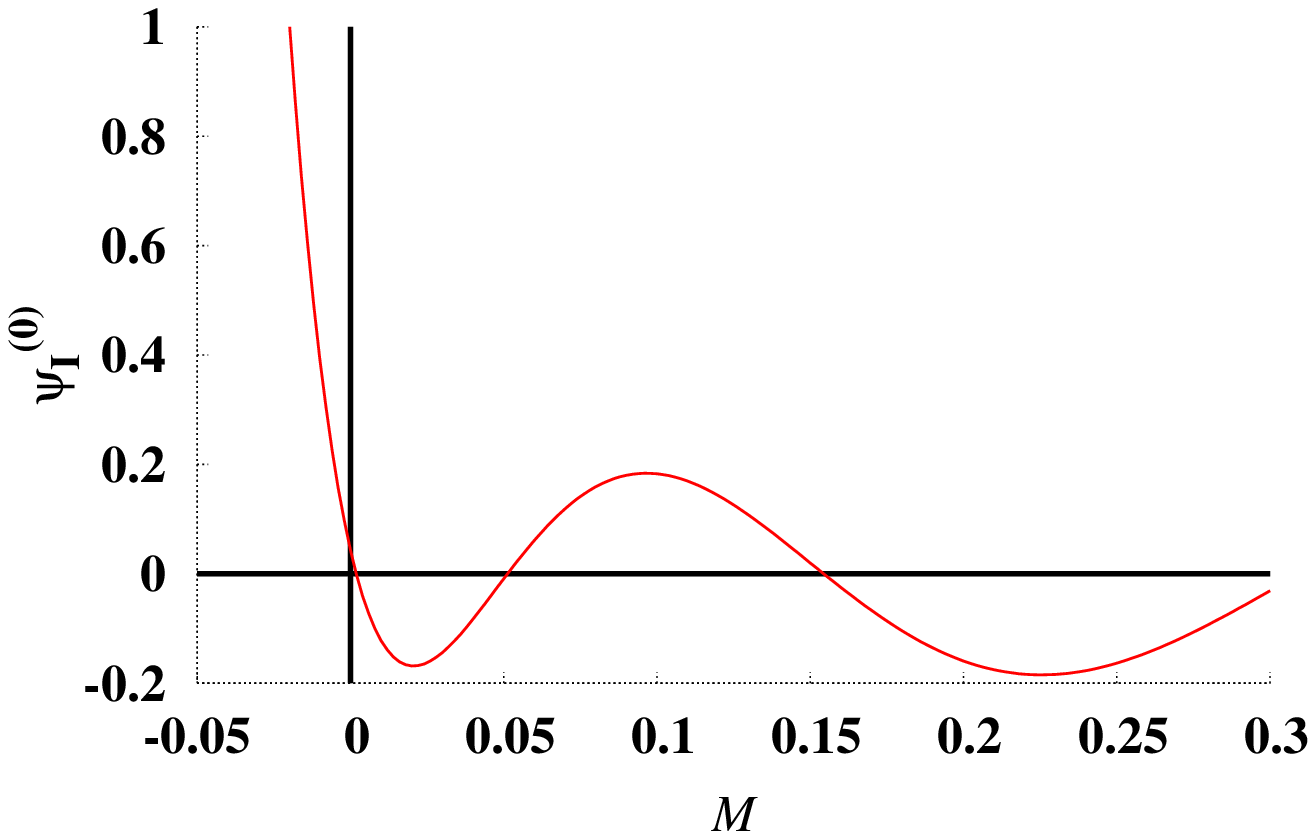}}}\quad
\subfigure[]{\resizebox{!}{0.2\linewidth}{\includegraphics{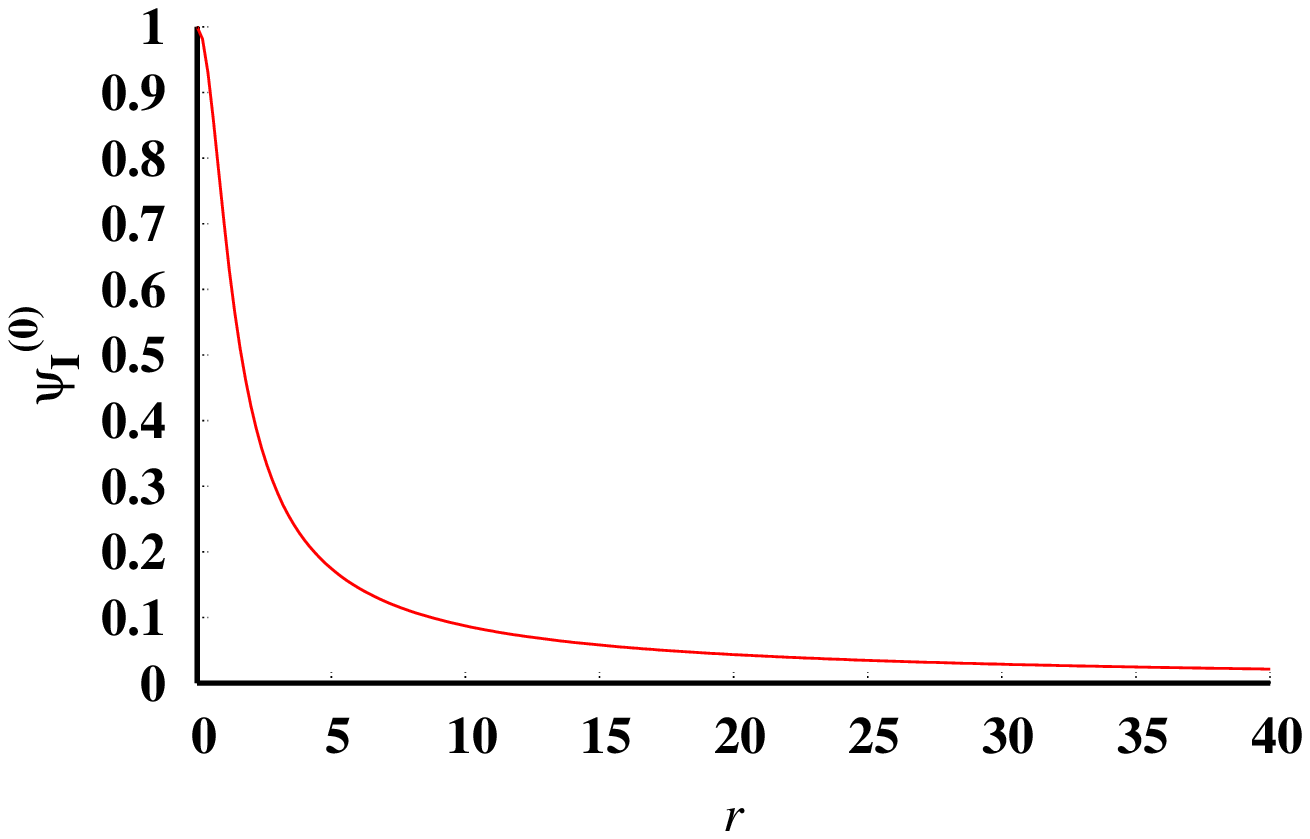}}}}
\end{center}
\vspace{-0.7cm}
\caption{\footnotesize Numerical results in the BPS case: $e =2$,
 $g=1$, $\xi=2$, $\eta =\mu =0$. (a) The potential $V^{(0)}_I$ has a
 zero energy fluctuation. 
 (b) The value of $\psi_I^{(0)}$ at a large distance ($r=20$) as
 function of $M$. 
 (c) The zero energy wave function $\psi_I^{(0)}$ as function of $r$. } 
\label{BPSpotentials}
\end{figure}

From a well-known theorem in one-dimensional quantum mechanics, it
follows that (valid for the lowest eigenvalues)
\begin{equation}
\left(M^{(k)}_I,M^{(k)}_{II}\right) \ge 
\left( \min\left[M^{(0)}_I,M^{(1)}_I\right],M^{(0)}_{II}\right)\ ,
\quad
\left(\tilde M^{(k)}_I,\tilde M^{(k)}_{II}\right) \ge 
\left( \min\left[\tilde M^{(0)}_I,\tilde M^{(1)}_I\right],
\tilde M^{(0)}_{II}\right)\ .
\end{equation}

\noindent
Hence, in order to exclude the existence of negative eigenvalues, it
suffices to study the six potentials: 
$V^{(0,1)}_I$, $\tilde V^{(0,1)}_I$, $V^{(0)}_{II}$ and $\tilde V^{(0)}_{II}$.

Fig.~\ref{potentials} shows that only the potentials $V^{(0)}_I$ and
$V^{(0)}_{II}$ can be sufficiently negative, such that one could expect 
negative eigenvalues. The potentials $V^{(1)}_I$ and $\tilde V^{(1)}_I$  
are even divergent at the core of the vortex.  Indeed, the results of
 the numerical analysis show (up to the numerical precision) that
 there are no negative eigenvalues for any of the 
potentials,  $V^{0,1}_I$, $\tilde V^{0,1}_I$, $V^0_{II}$ and $\tilde V^0_{II}$. 
We have checked this statement in a very wide range of values of the
couplings of the theory. This strongly suggests the absence of negative 
eigenvalues for all values of the couplings. The result is shown in
Fig.~\ref{NBPSpotentials} for a particular choice of parameters.  Note that all the potentials go to zero at large distance. This
means that there will be a set of positive eigenvalues that represent
the continuum states, which can be interpreted as interaction modes
with the massless Goldstone bosons of the bulk theory.

The status of the zero energy wave functions however is subtler. 
The wave functions for the non-BPS (Fig.~\ref{NBPSpotentials})
and the BPS cases (Fig.~\ref{BPSpotentials}) are both non-normalizable, hence they
should both be
interpreted as a part of the 
continuum.  
However, there is an important difference between the two
cases. In the BPS case, the wave function is limited and goes to zero
at infinity; its squared
norm diverges only logarithmically with the transverse volume 
($\sim\log L^2$).  In the
non-BPS case, the wave function is  unlimited, and it diverges linearly
(Fig.~\ref{NBPSpotentials}). 
This implies a divergence of its squared
norm being quadratic in the transverse volume ($\sim L^4$). 
Such a fluctuation changes the vacuum expectation values
of the scalars at infinity, thus does not correspond to a size zero mode (collective coordinate).  
The situation is clearer when our 
vortex system is put into a finite volume.  In the BPS case, 
the wave function of the semi-local mode asymptotically approaches
zero and is an 
(approximately) acceptable wave function.    In a box, it represents a
zero energy bound state.   On the contrary, 
the  wave function for the non-BPS case  is 
non-normalizable.   The corresponding zero energy state is eliminated and the
semi-local excitations have a mass gap. This observation strongly suggests that in the non-BPS case the size moduli disappear.

As a further check, we note that the potential which gives rise to the size zero modes in the BPS limit
(which we know exist in that case) is precisely  $V_I^{(0)}$.  
The zero
mode is thus given by a fluctuation of the field $\psi^{(0)}_I$. 
Using Eq.~(\ref{diagtransf}) we see that 
$\delta Q_{3I}=\delta \tilde Q_{3I} = \sqrt{2}\psi^{(0)}_I$, while all
other components of $Q_3$ and $\tilde Q_3$ vanish. This is exactly
the fluctuation we expect for the semi-local vortex, which in the BPS
limit is known to have the following form \cite{SYSemi,SemiL}: 
\beq
Q=\left(\begin{array}{ccc}
\phi_0(r) e^{ i \theta} & 0 & \chi (r)\\
0 & \phi_1(r) & 0\\
\end{array}
\right)\ . \label{semiloansatz}
\eeq

\section{Meta-stability versus Absolute Stability}
The results of the previous section indicate that the local (ANO-like) vortex 
embedded in a model with additional flavors is a local minimum of the
action. The problem whether or not this solution is a global 
minimum cannot be addressed with a fluctuation analysis only. In principle, the existence of an instability 
related to large fluctuations of the fields is still not excluded. 

However, we have a reason to believe that the local vortex
is truly stable, being the global minimum of the action.
Note that a vortex cannot dilute in the whole space. In fact,
such diluted configuration would have an energy that is equal to the
BPS bound, $T_\infty = 2\pi\xi = T_{\rm BPS}$, while the tension of the
local vortices, in the class of theories we are considering, is always
less: $T < T_{\rm BPS}$, as already noted in Section~\ref{model}.  To see
this, we will first show that the semi-local vortex solution, obtained
in the BPS limit, {\it is an approximate solution to the non-BPS
equations}, in the limit of an infinite size of the vortex. Then, it
is easy to check that the tension of such approximate configuration
converges to the BPS value. 
Let us write the equations of motion for the semi-local configuration
(\ref{semiloansatz}) 
\begin{align}
f_0''-\frac{f_0'}{r} &=e ^2
\left(f_1 (\phi_0^2-\phi_1^2+\chi ^2) +
f_0 (\phi_0^2+\phi_1^2+\chi ^2)-2 \chi ^2 \right)\ ,
\nonumber\\
f_1''-\frac{f_1'}{r} &=
g^2\left(f_1 (\phi_0^2+\phi_1^2+\chi ^2) +
f_0 (\phi_0^2-\phi_1^2+\chi ^2)-2\chi ^2 \right)\ , 
\nonumber\\
\phi_0''+ \frac{\phi_0'}{r} - \frac{(f_0+f_1)^2\phi_0}{4 r^2}
&=\frac{\phi_0}{2}
\left((\lambda_0+\lambda_1)^2+e ^2 A + g^2 B \right)\ , 
\nonumber\\
\phi_1''+\frac{\phi_1'}{r}-\frac{(f_0-f_1)^2\phi_1}{4 r^2} &=
\frac{\phi_1}{2}
\left((\lambda_0-\lambda_1)^2+e ^2 A - g^2 B \right)\ ,
\nonumber\\ 
\chi ''+\frac{\chi '}{r}-\frac{(f_0+f_1-2)^2\chi}{4 r^2} &=
\frac{\chi }{2}
\left((\lambda_0+\lambda_1)^2+e ^2 A+ g^2 B \right)\ , 
\nonumber\\
\lambda_0''+\frac{\lambda_0'}{r} &=
e ^2\left((\lambda_0+\lambda_1) (\phi_0^2+\chi ^2) + 
(\lambda_0-\lambda_1) \phi_1^2 
+ e ^2 \eta \, A \right)\ ,
\nonumber\\
\lambda_1''+\frac{\lambda_1'}{r} &= 
g^2\left((\lambda_0+\lambda_1) (\phi_0^2+\chi ^2) 
- (\lambda_0-\lambda_1) \phi_1^2 + g^2 \mu \, B \right) \ ,
\label{2ordinesemiloc}
\end{align}
where we have defined
\beq A \equiv \phi_0^2+\phi_1^2+\chi ^2-\xi+2 \eta  \lambda_0\ ,\quad
B \equiv \phi_0^2-\phi_1^2+\chi ^2+2 \mu \lambda_1\ . \eeq
For simplicity, we take the gauge couplings to be equal: $e = g$. 
Let us consider the following set of approximate solutions
corresponding to a semi-local BPS vortex in the limit of large size 
(lump limit): $a\gg1/e \sqrt \xi$: 
\beq
\left(\phi_0, \phi_1 , \chi \right)
= \sqrt{\frac{\xi}{2}} 
\left(\frac{r}{\sqrt{r^2+|a|^2}}, 1
,\frac{a}{\sqrt{r^2+|a|^2}}\right)\ ,
\quad
f_0 = f_1 = \frac{|a|^2}{r^2+|a|^2}\ , \quad 
\lambda_0 = \lambda_1 = 0\ .
\label{lump}
\eeq
The right-hand sides of Eqs.~(\ref{2ordinesemiloc}) calculated 
with this configuration vanish obviously.
The left-hand sides are derivative terms which disappear in the large
size limit: the set of fields above, trivially satisfy the fourth
and the last two equations, while they satisfy the other equations up
to terms of order $\mathcal{O}(1/|a|^2)$
\beq
\left| f_{0,1}''-\frac{f_{0,1}'}{r}\right| &=& 
\frac{8 |a|^2 r^2}{(r^2+|a|^2)^3}\leq \frac{32}{27 |a|^2}\ ,\nonumber\\
\left|\phi_0''+ \frac{\phi_0'}{r} - \frac{\phi_0(f_0+f_1)^2}{4 r^2}\right|
&=& \sqrt{\frac{\xi}{2}} \frac{2|a|^2 r}{(|a|^2+r^2)^{5/2}} \le 
\sqrt{\frac{\xi}{2}} 
\left(\frac{4}{5}\right)^{\frac{5}{2}}
\frac{1}{|a|^2}\ ,\nonumber\\
\left|\chi ''+\frac{\chi '}{r}-\frac{\chi (-2+f_0+f_1)^2}{4 r^2}\right|
&=& \sqrt{\frac{\xi}{2}} \frac{2|a|^3}{(|a|^2+r^2)^{5/2}} \le 
\sqrt{\frac{\xi}{2}} \frac{2}{|a|^2}\ .\nonumber
\label{2ordonlump}
\eeq
Thus, we see that the configuration of Eq.~(\ref{lump}), in the limit
$|a| \rightarrow \infty$, is a solution to the non-BPS equations of 
motion:  it represents an infinitely diluted vortex, having the
tension equal to the BPS value. 

We thus see that an infinitely wide vortex is not
favored energetically. The only remaining possibility is the
existence of a true minimum for a vortex with some intermediate size,
which involves also a non-trivial configuration for the semi-local
fields. This case is very unlikely. In fact, we could not find any such
solution in our numerical surveys. The relaxation method should have
detected such an unexpected stable vortex.

We consider our results as a strong evidence 
for the fact that the local vortex is a true minimum, 
absolutely stable against becoming a semi-local vortex.

\section{Low-energy Effective Theory}

In this short section we only make some considerations on the
low-energy effective theory of the vortex which follow from the
results of the previous sections.  

In the semi-local case, the vacuum has no mass gap, thus 
there are massless Nambu-Goldstone bosons in the bulk. 
This fact gives rise to subtleties (which we encountered in
Sec.~\ref{sec:zma}) when defining an effective theory 
for the vortex only. 
This is because both the vortex excitations and the massless
particles in the bulk appear as light modes.
As previously mentioned,
semi-local excitations are non-normalizable and the
corresponding states must be considered as bulk excitations.
In a rigorous effective theory, we should take into account only
localizable and normalizable modes, which discards the bulk
excitations\footnote{Similar issues have been already
 discussed for monopoles and vortices, see for example
 Refs. \cite{Tong:2008qd,Manton:1981mp,Weinberg:1979zt}.}. 
This holds both in the BPS and non-BPS cases. 
In fact, effective actions which include semi-local excitations have
already been proposed in the literature. The effective actions for BPS
semi-local vortices derived in Ref.~\cite{SYSemi,Eto:2006pg}, however,
are valid only in a finite volume. 

It seems to us that the following point has not been stressed clearly enough 
in the literature. We consider all the excitations related to
non-normalizable zero energy wave functions as part of the massless states of the bulk
theory. 
In the BPS case, the surviving supersymmetry enables us to describe
these fluctuations as collective coordinates of the vortex. It is known, for
example, that a single non-Abelian BPS vortex has non-normalizable
size parameters (in the case $N_{f}>N_{c}$). From the point of view
that we propose here, however, the effective action on the local
vortex is given by the orientational part only (normalizable
states). The other collective coordinates (size) give rise to zero energy fluctuations
 which represent bulk excitations. 

While these considerations might just be a matter of interpretation in
the BPS case, they could be important if one wants 
to push further the study of the non-BPS vortices we started in this 
work.

\section{Conclusion}

In this paper we have examined the stability of the non-Abelian
vortices in the context of an $\mathcal{N}=2$, $H=U(\Nc)$ model with
$\Nf > \Nc$ flavors and an $\mathcal{N}=1$ perturbation. The
particular perturbation chosen (the adjoint scalar mass terms) 
makes our vortices non-BPS. Local (ANO-like) vortices are found to
be stable; the vortex moduli corresponding to the semi-local vortices
(present in the BPS case) disappear, leaving intact the orientational
moduli -- $\mc P^{\Nc-1}$ in this particular model -- related to the
exact symmetry of the system.  

This conclusion seems to be very reasonable, as the narrow, ANO-like
vortices are needed to eliminate the regular monopoles from the
spectrum of the full system, in the  case our $H$ gauge theory arises
as a low-energy approximation of an underlying $G$ theory, after the
symmetry breaking $G \to H$. 
If the semi-local vortex would have survived the non-BPS
perturbations, the magnetic monopoles would be deconfined
\cite{SYSemi}.
Happily, this is not the case;
the Higgs phase of the (electric) low-energy theory
is actually a magnetic confinement phase.

\section*{Acknowledgments}
The authors thank David Tong for fruitful discussions.
S.~B.~G.~and W.~V.~are grateful to Department of Physics, Keio
University and Department of Physics, Tokyo Institute
of Technology for warm hospitality where part of this work has been
done.  The work of M.E.~is supported by the Research Fellowships of the Japan
Society for the Promotion of Science for Research Abroad.

\appendix
\section{Extension to $U(\Nc)$}\label{app}

In this appendix, we will extend the analysis of the vortices in
$SU(2) \times U(1)/{\mathbb Z}_2$ theory to the  $SU(\Nc)\times U(1)/{\mathbb Z}_{\Nc}$ with the $\Nf
\ge \Nc$ squarks in the fundamental representation.

The ${\cal N}=2$ multiplets are the following. The $SU(\Nc)$ vector
multiplet $(A_\mu, a=a_1+ia_2)$, the $U(1)$ vector multiplet
$(A_{0\mu},a_0=a_{0,1}+ia_{0,2})$ and the hypermultiplets 
$(Q,\tilde Q)$ in the fundamental representation of $SU(\Nc)$.
The squark fields $Q$ and $\tilde Q^\dagger$ are $\Nc \times \Nf$
matrices and their $U(1)$ charges are $+1/2$ and $-1/2$, respectively. 
The covariant derivatives are defined as
\beq
\D_\mu (Q,\tilde Q^\dagger) = 
(\p_\mu - iA_\mu - iA_{0\mu}/2) (Q,\tilde Q^\dagger)\ ,\quad
\D_\mu a = \p_\mu a - i[A_\mu,a]\ .
\eeq
We partially break SUSY by adding mass terms for the adjoint scalars
$a$ and $a_0$ in the following.
The bosonic part of the softly broken ${\cal N}=2$ Lagrangian 
${\cal L} = K - V$ takes the form
\beq
K &=& \Tr \left[ - \frac{1}{2g^2} F_{\mu\nu}F^{\mu\nu} + \frac{2}{g^2} \D_\mu a_i\D^\mu a_i
+ \D_\mu Q (\D^\mu Q)^\dagger + \D_\mu \tilde Q^\dagger (\D^\mu \tilde Q^\dagger )^\dagger
\right]\nonumber\\
&-& \frac{1}{4e^2} F_{0\mu\nu}F_0^{\mu\nu} + \frac{1}{e^2} \p_\mu
a_{0,i}\p^\mu a_{0,i}\ ,\\
V &=&
\Tr\bigg[
-\frac{4}{g^2} [a_1,a_2]^2
+ \frac{g^2}{4}\left<QQ^\dagger - \tilde Q^\dagger \tilde Q \right>^2
\bigg] + \frac{e^2}{8}  \left(\Tr[QQ^\dagger - \tilde Q^\dagger \tilde Q]\right)^2\nonumber\\
&+& \!\!\!\! g^2\Tr\left[ \left<Q\tilde Q + 2 \mu (a_1+ia_2)\right>
\left<Q\tilde Q + 2 \mu (a_1+ia_2)\right>^\dagger\right]\nonumber\\
&+& \!\!\!\! \frac{e^2}{2} \left|\Tr[Q \tilde Q] - \frac{\Nc\xi}{2} + 2\eta (a_{0,1}+ia_{0,2}) \right|^2
\!\!\!+ 2 \Tr\!\!\!\left[
\left(QQ^\dagger + \tilde Q^\dagger \tilde Q\right) \!\!
\left(a_i + \frac{1}{2} a_{0,i}{\bf 1}_N\right)^2
\right]\ .\label{eq:lag_su(n)}
\eeq
Our normalization is $\Tr[T^aT^b] = \delta^{ab}/2$ for the generators
of $SU(\Nc)$. 
The space-time metric is taken as 
$\eta_{\mu\nu} = (+,-,-,-)$. 
The masses of the adjoint scalars $(a,a_0)$ are $\mu$ and $\eta$.
$e$ is the $U(1)$ gauge coupling and $g$ is the $SU(\Nc)$ gauge coupling.
$\left< X \right>$ for an $N\times N$ matrix $X$ stands for the
traceless part of $X$ 
\footnote{
Useful identities:
$\Tr\left[\left<X\right>\left<Y\right>\right] = \Tr[XY]-\frac{\Tr[X]\Tr[Y]}{N}{\bf 1}_N$,
$\left<X + Y\right> = \left<X\right> + \left<Y\right>$.
}:
$\left<X\right> \equiv X - \frac{\Tr[X]}{N}{\bf 1}_N$ and
$\Tr[\left<X\right>] = 0$. 
We will choose the following vacuum in which to construct vortex
solutions 
\beq
Q = \tilde Q^\dagger = \sqrt{\frac{\xi}{2}}\left({\bf 1}_{\Nc}\ , {\bf 0}\right),\quad
a = a_0 = 0\ .
\eeq

\subsection{The Vortex Equations}

Let us make an Ansatz for the minimal winding vortex solution.
First of all, let us take an $SU(\Nc)$ generator
\beq
{\bf E} \equiv \frac{\Nc-1}{\Nc}~{\rm diag}\left(1,-\frac{1}{\Nc-1},\cdots,-\frac{1}{\Nc-1}\right)
\in \mathfrak{su}(\Nc),\quad \Tr[{\bf E}^2] = \frac{\Nc-1}{\Nc}\ .
\eeq
We make the following diagonal Ansatz
\beq
Q = \tilde Q^\dagger = 
\left({\rm diag}\left(
e^{i\theta}\phi_0(r),\phi_1(r),\cdots,\phi_1(r)\right), {\bf 0}
\right)\ ,\\
A_{0i} = - \epsilon_{ij} \frac{2}{N} \frac{x_j}{r^2} (1 - f_0(r))\ ,\quad
a_0 = a_{0,1} = \frac{2}{N}\lambda_0(r)\ ,\\
A_i = - \epsilon_{ij} \frac{x_j}{r^2} (1 - f_1(r)){\bf E}\ ,\quad
a = a_1 = \lambda_1(r) {\bf E}\ .
\eeq
Note $\left<Q\tilde Q + 2 \mu a\right> = \left<Q Q^\dagger + 2\mu a_1\right> = (\phi_0^2 - \phi_1^2+2\mu\lambda_1){\bf E}$
holds.

Inserting this Ansatz into the Lagrangian above leaves us with an
effective Lagrangian for the fields 
$f_0,f_1,\lambda_0,\lambda_1,\phi_0$ and $\phi_1$:
\beq
\tilde {\cal L} &=& 2\pi r (\tilde K - \tilde V)\ ,\\
\tilde K &=&
- \frac{\Nc-1}{\Nc g^2} \frac{f_1'{}^2}{r^2} - \frac{2(\Nc-1)}{\Nc g^2}\lambda_1'{}^2
- \frac{2}{\Nc^2e^2} \frac{f_0'{}^2}{r^2} - \frac{4}{\Nc^2 e^2} \lambda_0'{}^2
- 2\phi_0'{}^2  - 2(\Nc-1) \phi_1'{}^2\ ,\\
\tilde V &=& \frac{(\Nc-1)g^2}{\Nc} B^2
+ \frac{e^2}{2} A^2 
+ \frac{4}{\Nc^2} \phi_0^2 \left( \lambda_0 + (\Nc-1) \lambda_1 \right)^2 +
\frac{4(\Nc-1)}{\Nc^2} \phi_1^2 \left( \lambda_0 - \lambda_1\right)^2\nonumber\\
&&+~2 \frac{(f_0+(\Nc-1)f_1)^2\phi_0^2}{\Nc^2r^2} + 2 (\Nc-1)
\frac{(f_0-f_1)^2\phi_1^2}{\Nc^2r^2}\ .
\eeq
where we have defined
\beq
A &=& \phi_0^2 + (\Nc-1) \phi_1^2 - \frac{\Nc}{2} \xi + \frac{4}{\Nc}
\eta \lambda_0 \ ,\\
B &=& \phi_0^2 - \phi_1^2 + 2\mu \lambda_1\ .
\eeq
The vortex equations are simply the Euler-Lagrange equations for
$\tilde{\cal L}$: 
\beq
&&\!\!\!\!\!\!\!\!\!\!\!\!\!\!\!\!\!\!\!\!\!\!\!\!
f_0'' - \frac{f_0'}{r} 
- e^2 \left[  (f_0+(\Nc-1)f_1)\phi_0^2 +
 (\Nc-1)(f_0-f_1)\phi_1^2\right]=0 \ , \label{eq:eom_01}\\
&&\!\!\!\!\!\!\!\!\!\!\!\!\!\!\!\!\!\!\!\!\!\!\!\!
f_1'' - \frac{f_1'}{r} - \frac{2 g^2}{\Nc} \left[
(f_0+(\Nc-1)f_1)\phi_0^2 - (f_0-f_1)^2\phi_1^2
\right]
=0 \ ,\\
&&\!\!\!\!\!\!\!\!\!\!\!\!\!\!\!\!\!\!\!\!\!\!\!\!
\lambda_0'' + \frac{\lambda_0'}{r} - e^2 \bigg[
\frac{\Nc \eta e^2}{2} A
+ \left( \lambda_0 + (\Nc-1) \lambda_1 \right) \phi_0^2 +
(\Nc-1) \left( \lambda_0 - \lambda_1\right) \phi_1^2 \bigg] = 0\ ,\\
&&\!\!\!\!\!\!\!\!\!\!\!\!\!\!\!\!\!\!\!\!\!\!\!\!
\lambda_1'' + \frac{\lambda_1'}{r} - g^2\bigg[
\mu g^2 B + \frac{2}{\Nc} \left( \lambda_0 + (\Nc-1) \lambda_1 \right) \phi_0^2 -
\frac{2}{\Nc} \left( \lambda_0 - \lambda_1\right) \phi_1^2
\bigg] = 0\ ,\\
&&\!\!\!\!\!\!\!\!\!\!\!\!\!\!\!\!\!\!\!\!\!\!\!\!
\phi_0'' + \frac{\phi_0'}{r} - \bigg[
\frac{(\Nc-1)g^2}{\Nc} B
+ \frac{e^2}{2} A + \frac{2}{\Nc^2} \left( \lambda_0 + (\Nc-1) \lambda_1 \right)^2 
+ \frac{(f_0+(\Nc-1)f_1)^2}{\Nc^2r^2}\bigg]\phi_0 = 0\ ,\\
&&\!\!\!\!\!\!\!\!\!\!\!\!\!\!\!\!\!\!\!\!\!\!\!\!
\phi_1'' + \frac{\phi_1'}{r} - \bigg[
- \frac{g^2}{\Nc} B
+ \frac{e^2}{2} A + \frac{2}{\Nc^2} \left( \lambda_0 - \lambda_1\right)^2
+ \frac{(f_0-f_1)^2}{\Nc^2r^2}
\bigg] \phi_1 = 0\ .\label{eq:eom_06}
\eeq
These equations should be solved with the boundary conditions
\beq
\left(f_0,f_1,\lambda_0,\lambda_1,\phi_0,\phi_1\right) &\to&
\left(0,0,0,0,\sqrt{\frac{\xi}{2}}, \sqrt{\frac{\xi}{2}}\right)\ , 
\qquad {\rm as}\quad r \to \infty\ ,\\
\left(f_0,f_1,\lambda_0,\lambda_1,\phi_0,\phi_1\right) &\to&
\left(1,1,{\cal O}(1), {\cal O}(1), 0, {\cal O}(1)\right)\ , 
\qquad {\rm as}\quad r \to 0\ .
\eeq
All other solutions can be generated from this Ansatz by a flavor
rotation.

When we turn off the parameters $\mu$ and $\eta$, the model recovers
full ${\cal N}=2$ SUSY and the vortices therein become BPS states. One
of the common properties for various BPS states is 
that the EoMs can be reduced to first order differential equations. 
The energy density can be rewritten as follows
\beq
\frac{\tilde {\cal E}_{\rm BPS} }{2\pi r}
&=& \frac{\Nc-1}{\Nc g^2}\left[ \frac{f_1'}{r} - g^2 \left(\phi_0^2 - \phi_1^2 \right) \right]^2
+ \frac{2}{\Nc^2e^2} \left[ \frac{f_0'}{r} 
- \frac{\Nc e^2}{2} \left( \phi_0^2 + (\Nc-1) \phi_1^2 - \frac{\Nc}{2}\xi \right) \right]^2
\nonumber\\
&&+~ 2 \left[ \phi_0' - \frac{f_0+(\Nc-1)f_1}{\Nc r} \phi_0 \right]^2 + 2 (\Nc-1)
\left[ \phi_1' - \frac{f_0-f_1}{\Nc r} \phi_1 \right]^2\nonumber \\
&&-~ \xi \frac{f_0'}{r} + {\rm surface\ terms}\ , \nonumber\\
&\ge& - \xi \frac{f_0'}{r} \ ,
\eeq
where we have discarded $\lambda_0$ and $\lambda_1$ because of $\mu=\eta=0$.
The bound from below is saturated for the solutions satisfying the BPS
equations 
\beq
\frac{f_1'}{r} = g^2 \left(\phi_0^2 - \phi_1^2 \right)\ ,\quad
\frac{f_0'}{r} = \frac{\Nc e^2}{2} \left( \phi_0^2 + (\Nc-1) \phi_1^2
- \frac{\Nc}{2}\xi\right)\ ,
\label{eq:bps01}\\
\phi_0' = \frac{f_0+(\Nc-1)f_1}{\Nc r} \phi_0\ ,\quad
\phi_1' = \frac{f_0-f_1}{\Nc r} \phi_1\ .\qquad\qquad
\label{eq:bps02}
\eeq
The tension of the BPS vortex is
\beq
T = \int_0^\infty dr\ \tilde {\cal E}_{\rm BPS}=
2\pi\big[-f'_0(r)\big]^\infty_0 = 2\pi \xi\ .
\eeq

\subsection{The Perturbations}

In order to study whether the vortex solution in the previous
subsection is stable, we perturb the fields around the solution
considered to be the background. Let us denote the squark fields by 
\beq
Q = (Q_{\rm b} + \delta Q_1,\delta Q_2)\ ,\quad
\tilde Q^\dagger = ( Q_{\rm b} + \delta \tilde Q_1^\dagger, \delta
\tilde Q_2^\dagger)\ .
\eeq
Note $\delta Q_1,\delta \tilde Q_1$ are $\Nc\times\Nc$ and $\delta
Q_2,\delta\tilde Q_2$ are 
$\Nc \times (\Nf - \Nc)$. Since the last two are perturbations
around a vanishing background, they are completely decoupled
from all the other fields in the quadratic Lagrangian as we have seen 
in the main body of this paper. To unravel the mixed terms, we
make the following redefinition of fields 
$\delta Q_2 = (q + \tilde q^\dagger)/\sqrt2$ and $\delta \tilde Q_2 =
(q^\dagger - \tilde q)/\sqrt2$ as before. 
In order to derive the usual Schr\"odinger-type equations, we expand
$q,\tilde q$ as follows 
\beq
q = \sum_k e^{ik\theta} \left(
\begin{array}{c}
\psi_0^{(k)}(r)\\
\psi_1^{(k)}(r)\\
\vdots\\
\psi_{\Nc-1}^{(k)}(r)
\end{array}
\right)\ ,\quad
\tilde q = 
\sum_k e^{-ik\theta}
\left(
\tilde \psi_0^{(k)}(r),\ 
\tilde \psi_1^{(k)}(r),\cdots,
\tilde \psi_{\Nc-1}^{(k)}(r)
\right)\ ,
\eeq
where $\psi_i^{(k)}$ is an $\Nf-\Nc$ row vector and
$\tilde \psi_i^{(k)}$ is an $\Nf-\Nc$ column vector.

Substituting these into the Lagrangian (\ref{eq:lag_su(n)}), we obtain the
following four effective Lagrangians 
\beq
\frac{{\cal L}^{(k;0)}}{2\pi r} 
= - (\p_r \psi_{0,A}^{(k)})^2 - V_0^{(k)}(r) (\psi_{0,A}^{(k)})^2\ ,\qquad
\frac{{\cal L}^{(k;i)}}{2\pi r} 
= - (\p_r \psi_{i,A}^{(k)})^2 - V^{(k)}(r) (\psi_{i,A}^{(k)})^2\ ,\\
\frac{\tilde {\cal L}^{(k;0)}}{2\pi r} 
= - (\p_r \tilde \psi_{0,A}^{(0)})^2 - \tilde V_0^{(k)}(r) (\tilde
\psi_{0,A}^{(k)})^2\ ,\qquad
\frac{\tilde {\cal L}^{(k;i)}}{2\pi r} 
= - (\p_r \tilde \psi_{i,A}^{(0)})^2 - \tilde V^{(k)}(r) (\tilde
\psi_{i,A}^{(k)})^2\ ,
\eeq
where $A$ denotes the flavor index: $A=1,2,\cdots,\Nf-\Nc$ and
$i=1,\cdots,\Nc-1$. 
The Schr\"odinger potentials are defined as
\beq
&&\!\!\!\!\!\!\!\!\!\!\!\!\!\!\!\!\!\!\!\!\!\!\!\!
V_0^{(k)} = \frac{(\Nc-1)g^2}{\Nc}  B
+ \frac{e^2}{2} A +
\frac{2(\lambda_0 + (\Nc-1)\lambda_1)^2}{\Nc^2}
+\left(\frac{f_0 + (\Nc-1)f_1 + \Nc (k-1)}{r \Nc}\right)^2,\\
&&\!\!\!\!\!\!\!\!\!\!\!\!\!\!\!\!\!\!\!\!\!\!\!\!
V^{(k)} = - \frac{g^2}{\Nc}  B
+ \frac{e^2}{2} A +\frac{2(\lambda_0 - \lambda_1)^2}{\Nc^2}
+ \left(\frac{f_0 - f_1 + \Nc k}{r \Nc}\right)^2,\\
&&\!\!\!\!\!\!\!\!\!\!\!\!\!\!\!\!\!\!\!\!\!\!\!\!
\tilde V_0^{(k)} = -\frac{(\Nc-1)g^2}{\Nc}  B
- \frac{e^2}{2} A + \frac{2(\lambda_0 + (\Nc-1)\lambda_1)^2}{\Nc^2}
+\left(\frac{f_0 + (\Nc-1)f_1 + \Nc (k-1)}{r \Nc}\right)^2,\\
&&\!\!\!\!\!\!\!\!\!\!\!\!\!\!\!\!\!\!\!\!\!\!\!\!
\tilde V^{(k)} =  \frac{g^2}{\Nc}  B
- \frac{e^2}{2} A + \frac{2(\lambda_0 - \lambda_1)^2}{\Nc^2}
+ \left(\frac{f_0 - f_1 + \Nc k}{r \Nc}\right)^2.
\eeq
As expected, we have obtained only four different Schr\"odinger-type 
equations independent of $\Nc,\Nf$ $(\Nc > 1,\Nf>\Nc)$. Therefore, the results we found in the
minimal example with $\Nc=2,\Nf=3$ in the body of this paper are 
valid more generally.  
Also, when the gauge couplings are fine-tuned as $e^2/2 = g^2/\Nc$, the
$\Nc$-dependence  disappears from the equations altogether.


\begin{thebibliography}{99}

\bibitem{'tHooft:1981ht}
 G.~'t Hooft,
 Nucl.\ Phys.\  B {\bf 190}, 455 (1981);
 S.~Mandelstam,
 Phys.\ Lett.\  B {\bf 53}, 476 (1975).

\bibitem{HT}
 A.~Hanany and D.~Tong,
 JHEP {\bf 0307} (2003) 037
 [arXiv:hep-th/0306150].


\bibitem{ABEKY}
R.~Auzzi, S.~Bolognesi, J.~Evslin, K.~Konishi and A.~Yung,
Nucl.\ Phys.\ B {\bf 673} (2003) 187
[arXiv:hep-th/0307287].


\bibitem{ABEK}
 R.~Auzzi, S.~Bolognesi, J.~Evslin and K.~Konishi,
 Nucl.\ Phys.\  B {\bf 686} (2004) 119
 [arXiv:hep-th/0312233].

\bibitem{tmon}
 D.~Tong,
 Phys.\ Rev.\  D {\bf 69} (2004) 065003
 [arXiv:hep-th/0307302].


\bibitem{SY}  
  M.~Shifman and A.~Yung,
 Phys.\ Rev.\  D {\bf 70} (2004) 045004
 [arXiv:hep-th/0403149].

 \bibitem{HT2}
 A.~Hanany and D.~Tong,
 JHEP {\bf 0404} (2004) 066
 [arXiv:hep-th/0403158].





\bibitem{GSY}
 A.~Gorsky, M.~Shifman and A.~Yung,
 Phys.\ Rev.\  D {\bf 71} (2005) 045010
 [arXiv:hep-th/0412082].





\bibitem{Duality}
M. Eto, L. Ferretti,  K. Konishi, G. Marmorini, M. Nitta, K. Ohashi, W. Vinci,  N. Yokoi,
Nucl.Phys. {\bf B780} 161-187, 2007
  [arXiv: hep-th/0611313].

\bibitem{Baptista:2008ex}
  J.~M.~Baptista,
  arXiv:0810.3220 [hep-th].


\bibitem{Eto:2006pg}
M.~Eto, Y.~Isozumi, M.~Nitta, K.~Ohashi and N.~Sakai,
J.\ Phys.\ A {\bf 39} (2006) R315
[arXiv:hep-th/0602170].

\bibitem{Shifman:2007ce}
 M.~Shifman and A.~Yung,
 Rev.\ Mod.\ Phys.\  {\bf 79} (2007) 1139
 [arXiv:hep-th/0703267].


\bibitem{Tong:2005un}
 D.~Tong,
 arXiv:hep-th/0509216.


\bibitem{Tong:2008qd}
 D.~Tong,
 arXiv:0809.5060 [hep-th].


\bibitem{HashiTong}  
 K.~Hashimoto and D.~Tong,
 JCAP {\bf 0509} (2005) 004
 [arXiv:hep-th/0506022].

\bibitem{Eto:2005yh}
M.~Eto, Y.~Isozumi, M.~Nitta, K.~Ohashi and N.~Sakai,
Phys.\ Rev.\ Lett.\ {\bf 96} (2006) 161601
[arXiv:hep-th/0511088].

\bibitem{ASY}   R.~Auzzi, M.~Shifman and A.~Yung,
 Phys.\ Rev.\  D {\bf 73} (2006) 105012
 [Erratum-ibid.\  D {\bf 76} (2007) 109901]
 [arXiv:hep-th/0511150].


\bibitem{sevenHW}
 M.~Eto, K.~Konishi, G.~Marmorini, M.~Nitta, K.~Ohashi, W.~Vinci and N.~Yokoi,
 Phys.\ Rev.\  D {\bf 74} (2006) 065021
 [arXiv:hep-th/0607070].



\bibitem{SYSemi}
 M.~Shifman and A.~Yung,
 Phys.\ Rev.\  D {\bf 73} (2006) 125012
 [arXiv:hep-th/0603134].


\bibitem{SemiL}  
   M. Eto, J. Evslin, K. Konishi, G. Marmorini, M. Nitta, K. Ohashi, W. Vinci,  N. Yokoi,   
 Phys.\ Rev.\  D {\bf 76} (2007) 105002
 [arXiv:0704.2218 [hep-th]].

\bibitem{Vachaspati:1991dz}
 T.~Vachaspati and A.~Achucarro,
 Phys.\ Rev.\  D {\bf 44}, 3067 (1991).

\bibitem{achurev}
 A.~Achucarro and T.~Vachaspati,
 Phys.\ Rept.\  {\bf 327} (2000) 347
 [Phys.\ Rept.\  {\bf 327} (2000) 427]
 [arXiv:hep-ph/9904229].


\bibitem{AEV1} 
 R.~Auzzi, M.~Eto and W.~Vinci,
 JHEP {\bf 0711} (2007) 090
 [arXiv:0709.1910 [hep-th]].


\bibitem{AEV2} 
 R.~Auzzi, M.~Eto and W.~Vinci,
 JHEP {\bf 0802} (2008) 100
 [arXiv:0711.0116 [hep-th]].
 M.~Eto,
 arXiv:0810.4895 [hep-th].



\bibitem{KF} 
 L.~Ferretti and K.~Konishi,
 arXiv:hep-th/0602252.


\bibitem{GFK}
 L.~Ferretti, S.~B.~Gudnason and K.~Konishi,
 Nucl.\ Phys.\  B {\bf 789} (2008) 84
 [arXiv:0706.3854 [hep-th]].




\bibitem{AGG}
 M.~Eto, T.~Fujimori, S.~B.~Gudnason, K.~Konishi, M.~Nitta, K.~Ohashi and W.~Vinci,
 arXiv:0802.1020 [hep-th].
 W.~Vinci,
 arXiv:0810.2449 [hep-th].

\bibitem{KonishiMN} 
 K.~Konishi,
 arXiv:0809.1374 [hep-th].


\bibitem{DKO}
 D.~Dorigoni, K.~Konishi and K.~Ohashi,
 arXiv:0801.3284 [hep-th].

\bibitem{heterotic}
 M.~Edalati and D.~Tong,
 JHEP {\bf 0705} (2007) 005
 [arXiv:hep-th/0703045];
 D.~Tong,
 JHEP {\bf 0709} (2007) 022
 [arXiv:hep-th/0703235];
 M.~Shifman and A.~Yung,
 Phys.\ Rev.\  D {\bf 77} (2008) 125016
 [arXiv:0803.0158 [hep-th]];
 M.~Shifman and A.~Yung,
 Phys.\ Rev.\  D {\bf 77} (2008) 125017
 [arXiv:0803.0698 [hep-th]];

\bibitem{Seiberg:1994bz}
 N.~Seiberg,
 Phys.\ Rev.\  D {\bf 49}, 6857 (1994)
 [arXiv:hep-th/9402044].

\bibitem{sdvort}
M.~Shifman and A.~Yung,
 Phys.\ Rev.\  D {\bf 76} (2007) 045005
 [arXiv:0705.3811 [hep-th]];
 S.~Bolognesi,
 arXiv:0807.2456 [hep-th].

\bibitem{n1st1}
 V.~Markov, A.~Marshakov and A.~Yung,
 Nucl.\ Phys.\  B {\bf 709} (2005) 267
 [arXiv:hep-th/0408235].

\bibitem{n1st2}
 R.~Auzzi and S.~P.~Kumar,
 arXiv:0810.3201 [hep-th].

\bibitem{ps}
 J.~Polchinski and M.~J.~Strassler,
 arXiv:hep-th/0003136.


\bibitem{thooft}
 G.~'t Hooft,
 Nucl.\ Phys.\  B {\bf 153} (1979) 141.

\bibitem{penin}
 A.~A.~Penin, V.~A.~Rubakov, P.~G.~Tinyakov and S.~V.~Troitsky,
 Phys.\ Lett.\  B {\bf 389} (1996) 13
 [arXiv:hep-ph/9609257].


\bibitem{Goddard:1976qe}
 P.~Goddard, J.~Nuyts and D.~I.~Olive,
 Nucl.\ Phys.\  B {\bf 125}, 1 (1977);
 F.~A.~Bais,
 Phys.\ Rev.\  D {\bf 18}, 1206 (1978);
 E.~J.~Weinberg,
 Nucl.\ Phys.\  B {\bf 167}, 500 (1980).




\bibitem{Eto:2007hf}
 M.~Eto, K.~Hashimoto and S.~Terashima,
 JHEP {\bf 0709}, 036 (2007)
 [arXiv:0706.2005 [hep-th]].




\bibitem{Hindmarsh:1991jq}
M.~Hindmarsh,
Phys.\ Rev.\ Lett.\ {\bf 68} (1992) 1263;  
 Nucl.\ Phys.\  B {\bf 392} (1993) 461
 [arXiv:hep-ph/9206229].



\bibitem{Vachasp}
A.~Achucarro, K.~Kuijken, L.~Perivolaropoulos and T.~Vachaspati,
Nucl.\ Phys.\  B {\bf 388} (1992) 435.


\bibitem{leese}
 R.~A.~Leese,
 Phys.\ Rev.\  D {\bf 46} (1992) 4677;
 R.~A.~Leese and T.~M.~Samols,
 Nucl.\ Phys.\  B {\bf 396} (1993) 639.


\bibitem{Abrikosov:1956sx}
A.~A.~Abrikosov, 
Sov.\ Phys.\ JETP {\bf 5}
(1957) 1174 [Zh.\ Eksp.\ Teor.\ Fiz.\ {\bf 32} (1957) 1442].


\bibitem{Nielsen:1973cs}
H.~B.~Nielsen and P.~Olesen,
Nucl.\ Phys.\ B {\bf 61} (1973) 45.


\bibitem{CKM}
 G.~Carlino, K.~Konishi and H.~Murayama,
 Nucl.\ Phys.\  B {\bf 590} (2000) 37
 [arXiv:hep-th/0005076].

\bibitem{Lindstrom:1983rt}
U.~Lindstr\"om and M.~Ro\v{c}ek,
Nucl.\ Phys.\ B {\bf 222} (1983) 285; 
 I.~Antoniadis and B.~Pioline,
 Int.\ J.\ Mod.\ Phys.\  A {\bf 12}, 4907 (1997)
 [arXiv:hep-th/9607058].

\bibitem{Vainshtein:2000hu}
 A.~I.~Vainshtein and A.~Yung,
 Nucl.\ Phys.\  B {\bf 614}, 3 (2001)
 [arXiv:hep-th/0012250].

\bibitem{Manton:1981mp}
 N.~S.~Manton,
 Phys.\ Lett.\  B {\bf 110}, 54 (1982).

\bibitem{Weinberg:1979zt}
 E.~J.~Weinberg,
 Nucl.\ Phys.\  B {\bf 167}, 500 (1980).


\end{thebibliography}
\end{document}